\documentclass[preprint,showpacs,preprintnumbers,amsmath,amssymb,aps,prd,floatfix,nofootinbib]{revtex4-1}
\usepackage[colorlinks=true,linkcolor=blue,filecolor=blue,urlcolor=blue,citecolor=blue,pdftex=true,plainpages=false]{hyperref}
\usepackage{amsmath,amssymb,amsbsy,amsfonts,bm}
\usepackage{graphicx}
\usepackage[mathscr]{euscript}
\usepackage{multirow}

\usepackage{graphicx}
\usepackage{subfigure}

\newcommand\Tau{\mathrm{T}}
\newcommand\Ol{\mathcal{O}}

\usepackage{color}

\def\rcite#1{Ref.~\cite{#1}}
\def\Rcite#1{Reference~\cite{#1}}

\def\eqn#1{\label{eq:#1}}

\def\eq#1{Eq.~(\ref{eq:#1})}

\def\eqs#1#2{Eqs.~(\ref{eq:#1}) and (\ref{eq:#2})}

\def\figref#1{Fig.~\ref{fig:#1}}
\def\Figref#1{Figure~\ref{fig:#1}}

\def\tabref#1{Table~\ref{tab:#1}}
\def\tabrefs#1#2{Tables~\ref{tab:#1} and \ref{tab:#2}}
\def\gtwid{{\,\raise.3ex\hbox{$>$\kern-.75em\lower1ex\hbox{$\sim$}}\,}}
\def\ltwid{{\,\raise.3ex\hbox{$<$\kern-.75em\lower1ex\hbox{$\sim$}}\,}}
\def\cO{{\cal O}}
\begin{document}


\title{Gradient flow and scale setting\\ on MILC HISQ ensembles}



\author{A.~Bazavov}
\altaffiliation[Present address:~]{Department of Physics and Astronomy, University of Iowa, Iowa City, Iowa, 52242 USA}
\affiliation{Physics Department, Brookhaven National Laboratory, Upton, New York 11973, USA}

\author{C.~Bernard}
\author{N.~Brown}
\email[]{brownnathan@wustl.edu}
\author{J.~Komijani}
\affiliation{Department of Physics, Washington University, St. Louis, Missouri 63130, USA}

\author{C.~DeTar}
\author{J.~Foley}
\author{L.~Levkova}
\affiliation{Department of Physics and Astronomy, University of Utah, Salt Lake City, Utah 84112, USA}

\author{Steven~Gottlieb}
\affiliation{Department of Physics, Indiana University, Bloomington, Indiana 47405, USA}

\author{U.M.~Heller}
\affiliation{American Physical Society, One Research Road, Ridge, New York 11961, USA}

\author{J.~Laiho}
\affiliation{Department of Physics, Syracuse University, Syracuse, New York 13244, USA}

\author{R.L.~Sugar}
\affiliation{Physics Department, University of California, Santa Barbara, California 93106, USA}

\author{D.~Toussaint}
\affiliation{Physics Department, University of Arizona Tucson, Arizona 85721, USA}

\author{R.S.~Van~de~Water}
\affiliation{Fermi National Accelerator Laboratory, Batavia, Illinois 60510, USA}

\collaboration{MILC Collaboration}

\date{\today}

\begin{abstract}
We report on a scale determination with gradient-flow techniques on the $N_f=2+1+1$ highly improved staggered quark ensembles generated by the MILC Collaboration. 
The ensembles include four lattice spacings, ranging from approximately 0.15 to 0.06 fm, and both  physical and unphysical values of the quark masses.
The scales $\sqrt{t_0}/a$ and $w_0/a$ and their tree-level improvements, $\sqrt{t_{0,{\rm imp}}}$ and $w_{0,{\rm imp}}$, are computed on each ensemble using Symanzik flow and the cloverleaf definition of the energy density $E$. 
Using a combination of continuum chiral-perturbation theory and a Taylor-series ansatz for the lattice-spacing and strong-coupling dependence, the results are simultaneously extrapolated to the continuum and interpolated to physical quark masses. 
We determine the scales $\sqrt{t_0} = 0.1416({}_{-5}^{+8})$ fm and $w_0 = 0.1714({}_{-12}^{+15})$ fm, where the errors are sums, in quadrature, of statistical and all systematic errors.  
The precision of $w_0$ and $\sqrt{t_0}$ is comparable to or more precise than the best previous estimates, respectively. 
We then find the continuum mass dependence of $\sqrt{t_0}$ and $w_0$, which will be useful for estimating the scales of new ensembles. 
We also estimate the integrated autocorrelation length of $\langle E(t) \rangle$. For long flow times, the autocorrelation length of $\langle E \rangle$ appears to be comparable to that of the topological charge.
\end{abstract}


\maketitle


\section{Introduction \label{intro}}
Scale setting holds central importance in lattice QCD for two reasons. 
First, the continuum extrapolation of any quantity, dimensionful or dimensionless, requires a precise determination of the relative scale between ensembles with different bare couplings. 
Second, the precision to which one may determine a dimensionful quantity in physical units is limited by the precision of the scale in physical units (the {\it absolute scale}). 
Because scale setting limits the precision of so many calculations, it is important to identify quantities with the highest level of precision to set the scale. 

To make progress towards this goal a thorough understanding of the restrictions on quantities that may be used for scale setting is required. 
In principle, any dimensionful quantity that is finite in the continuum limit may be employed. 
The relative scale may be set by calculating a dimensionful quantity and comparing its value in lattice units at different lattice spacings for the same quark masses. 
For absolute scale setting, one needs to compare the quantity in lattice units to the physical value. 
If the quantity is experimentally accessible the comparison to the physical value is straightforward. 
For a quantity that is inaccessible to experiments, its physical value in the continuum is inferred by comparison to an experimental quantity.
In other words, an experimental quantity may be used directly for relative and absolute scale setting, but a quantity that is inaccessible to experiments requires the lattice measurement of a second, experimentally accessible quantity for absolute scale setting. 
The use of a nonexperimental quantity for scale setting may still be worthwhile if it can be determined on the lattice with small statistical and systematic errors for relatively small computational cost. 
This is due to the large gain in control over continuum extrapolations at the cost of a small decrease in the precision of absolute scales.
This has led to the consideration of theoretically motivated, but not experimentally measurable, quantities such as $r_0$ and $r_1$ \cite{r0,r1}, $F_{p4s}$ \cite{hisq_ensembles}, and, more recently, $\sqrt{t_0}$ \cite{t0} and $w_0$ \cite{bmw} from gradient flow \cite{smoothing_origin,luscher_origin}. 

The ideal scale-setting quantity has small statistical and systematic errors. 
However, since systematic errors arise from a variety of sources, such as discretization effects, dependence on the simulation (possibly unphysical) quark masses, finite-volume effects, and excited states, it is difficult to reduce all error sources simultaneously. 
For example, the scales $r_0$ and $r_1$ are computed from asymptotic fits in time $t$ to the heavy-quark potential $V(r)$ with quark separation $r$, such that $r^2 dV/{dr} = 1.65$ or $1$, for $r=r_0$ and $r_1$, respectively \cite{r0,r1}. 
The statistical errors in $V(r)$ are generally small,  but they grow with $t/a$ and may become a
problem at small lattice spacings where larger values of $t/a$ are needed to reduce systematic errors from excited states \cite{hisq_ensembles}.  
As another example, consider $F_{p4s}$, the fictitious pseudoscalar decay constant with degenerate valence quarks of mass $m_{\rm v}=0.4m_s$ and physical sea-quark masses \cite{hisq_ensembles}. 
The value of the valence-quark mass is chosen to be heavy enough to make it not too expensive to compute the correlators, but light enough for chiral-perturbation theory to apply. 
However, $F_{p4s}$ has strong dependence on the valence-quark mass.  
Thus, relatively small errors in determining $am_s$, the physical value of the strange-quark mass in lattice units, may lead to significant errors in $aF_{p4s}$ through the value of the valence mass, $am_{\rm v}=0.4am_s$. 
Further, the required asymptotic fits to correlators  are difficult to automate and usually require significant human intervention.

Gradient flow \cite{smoothing_origin, luscher_origin} has received considerable attention \cite{t1, theory, renorm, thermo} over the past few years because it is a theoretically grounded smoothing operation that is simple to implement and can be used to obtain precisely determined scales. 
The basis for scale setting with gradient flow is the determination of the flow time for which a dimensionless, precise, and easily computable quantity is smoothed to a predefined value. 
The original quantity proposed by L\"uscher, $t_0$, is defined through the gauge field energy density \cite{t0}. 
Most modifications focus on reducing discretization errors in the same underlying flow or observable \cite{t1, bmw, tshift, treelevel}. 
All of these scales can be easily computed to a statistical  precision of $0.1\%$ or less and have small quark-mass dependence.
Finite-volume effects, the only remaining sources of systematic error for relative scale setting, may also be
kept very small.

Here, we present our computation of the gradient-flow scales $\sqrt{t_0}/a$ and $w_0/a$ on the MILC, (2+1+1)-flavor, highly improved staggered quark (HISQ) ensembles \cite{hisq_action, hisq_ensembles}. 
The HISQ configurations used in this analysis cover lattice spacings from $a\approx 0.15$ to $0.06$ fm and include ensembles with physical, or heavier than physical, light-quark masses,  and physical, or lighter than physical, strange-quark mass. 
The charm-quark mass is kept near its physical value. 
We perform a continuum extrapolation and interpolation to physical quark masses of $w_0F_{p4s}$ and $\sqrt{t_0}F_{p4s}$ to determine the two scales in physical units, using our previous determination of $F_{p4s}$ in physical units \cite{fp4s}. 
We find $\sqrt{t_0} = 0.1416({}_{-5}^{+8})$ fm and $w_0 = 0.1714({}_{-12}^{+15})$ fm, where statistical and all systematic errors have been added in quadrature.

We start with a review of the relevant theoretical details, including the gradient-flow equation in Sec.~\ref{gfe}, definitions of the scales $t_0$ and $w_0$ in Sec.~\ref{scales}, chiral-perturbation theory for flow quantities in Sec.~\ref{chipt}, and lattice-spacing dependence in Sec.~\ref{disc}. 
The computational setup is described in Sec.~\ref{setup}. 
We discuss the raw lattice results in Sec.~\ref{data}, include a brief comparison of the results for different ensemble-generation algorithms in Sec.~\ref{rhmc}, and estimate the integrated autocorrelation lengths in Sec.~\ref{auto}.
Leading-order adjustments for charm-quark-mass mistuning are performed in Sec.~\ref{charm}, and a simple extrapolation to the continuum of the results on the physical-mass ensembles is presented in Sec.~\ref{naive}. 
Section~\ref{regress} then describes the quark-mass interpolation and continuum extrapolation.
We present our results for $w_0$ and $\sqrt{t_0}$ in physical units in Sec.~\ref{cont}, and include comparisons with our earlier preliminary results. 
The continuum mass dependence of $\sqrt{t_0}$ and $w_0$ is deduced from our fits in Sec.~\ref{mass} and used to compare the scales determined from the gradient flow to those determined from $F_{p4s}$ in \rcite{fp4s}; knowing the continuum mass dependence will be useful in determining the scales of new ensembles.  
Section~\ref{end} compares our results to those of other collaborations, and tabulates the precision of various methods for relative scale setting.

Preliminary versions of this analysis have been described in Refs. \cite{prelim1} and \cite{prelim2}.

\section{Review of Gradient Flow \label{theory}}
This section summarizes the theoretical details of gradient flow from Refs.~\cite{smoothing_origin,luscher_origin,t0,bmw,treelevel,chipt} that are relevant to the scale-setting analysis in later sections.

\subsection{Diffusion equation \label{gfe}}
Gradient flow \cite{smoothing_origin, luscher_origin} is a smoothing of the original gauge fields $A$ towards stationary points of the action $S$. 
The new, smoothed gauge fields $B(t)$ are functions of the ``flow time" $t$ and are updated according to the diffusionlike equation below,
where $g_0$ is the bare coupling. 
\begin{eqnarray} 
	&&\frac{d B_\mu}{dt}  = - g_0^2\frac{\partial S}{\partial B_\mu} = D_\nu G_{\nu\mu}\,, \hspace{3mm} 
		B_\mu(0) = A_\mu\,, \\
	&&D_\nu X = \partial_\nu X + \left[B_\nu,\, X \right] \,, \hspace{3mm}
	   G_{\nu\mu} = \partial_\nu B_\mu - \partial_\mu B_\nu + \left[ B_\nu,\, B_\mu \right]\,. \nonumber
\end{eqnarray}
				
On the lattice, the Yang-Mills action is replaced by an appropriate discretized version. 
The gauge link $V(t)_{i,\mu}$ at site $i$ in direction $\mu$ is updated in time according to
\begin{equation}
\frac{d V(t)_{i,\mu}}{dt}  = - g_0^2 \frac{\partial S(V)}{\partial V_{i,\mu}} V_{i,\mu} \,, \hspace{3mm} V_{i,\mu}(0) = U_{i,\mu}
\end{equation}
	
\noindent
The change of $V(t)$ with flow time explicitly follows the steepest descent of the action with respect to the gauge field, with an additional factor of $V_{i,\mu}$ in the lattice formulation to ensure gauge covariance. 
For more details on the SU(3)-valued derivative, see the Appendix of Ref.~\cite{t0}.

As the flow time $t$ increases, the gauge fields diffuse and short-distance lattice artifacts are removed. 
After modifying the flow equation with a flow-time-dependent gauge transformation of the field one can explicitly see the suppression of high momenta in the leading-order perturbative expansion of the gauge field in powers of the coupling $g_0$ \cite{t0}:
\begin{equation} 
B_\mu(x,t)\approx \frac{1}{(4\pi t)^2} \int\,d^4y A_\mu(y) e^{-(x-y)^2/(4t)} \ , \hspace{5mm}
\tilde{B}_\mu(p,t) \approx \tilde{A}_\mu(p) e^{-tp^2} \ .
\end{equation}

\subsection{Gradient-flow scales \label{scales}}
The process of gradient flow introduces a dimensionful, independent variable, the flow time. 
Since all quantities calculated from smoothed gauge links will be functions of the flow time, one may define a scale by choosing a reference time at which a chosen dimensionless quantity reaches a predefined value. 
If the dimensionless quantity is also finite in the continuum limit, then the reference time scale will be independent of the lattice spacing up to discretization corrections in powers of $a^2$. 
One of the easiest dimensionless quantities to calculate with only gauge fields is the average total energy within a smoothed volume $V\propto t^2$.  
Up to a dimensionless constant, this is equivalent to calculating the product of the energy density and squared flow time $t^2 \langle E(t) \rangle$. 
L\"uscher and Weisz have shown that the energy density is finite to all orders (when expressed in terms of renormalized quantities) \cite{all_order}, so $t^2 \langle E(t) \rangle$ is a suitable candidate for setting the scale. 
A fiducial point $c$ is chosen, and the reference scale is defined to be the flow time $t_0$ where
\begin{equation}\eqn{t0_definition}
t_0^2 \langle E(t_0) \rangle = c\ .
\end{equation}

\noindent
The fiducial point should be chosen so that for simulated lattice spacings $a$ and volumes $V=L^3T$ (with $T\ge L$), the reference time scale $t_0$ falls within $a \ll \sqrt{8t_0} \ll aL$.  
The value of $c=0.3$ has been found, empirically, to satisfy this relation \cite{t0, bmw}.  
A larger fiducial point of $c=2/3$ has also been proposed in order to reduce discretization errors, at the expense of somewhat larger finite-volume effects \cite{t1}.

The renormalized expansion of $\langle E(t) \rangle$ to second order in $g$ shows $t^2 \langle E(t) \rangle$ is approximately constant \cite{t0}. 
For small flow times this agrees with computational results, but for larger flow times (including the scale $t_0$) $t^2 \langle E(t) \rangle$ is found empirically to be linear in $t$ \cite{t0,bmw}. 
The transition of $\langle E(t) \rangle$ from $t^{-2}$ to $t^{-1}$ dependence is nonperturbative. 
However, we expect discretization errors to enter primarily for small flow times, before the lattice details are smoothed away.  
In accordance with this expectation, empirical evidence suggests that discretization effects have less impact on the slope of  $t^2 \langle E(t) \rangle$ at comparatively larger flow times near the fiducial point, than they do on $t^2 \langle E(t) \rangle$ itself \cite{bmw}. 
Assuming the property is general, an improvement to the scale $t_0$ is computed by considering the slope:
\begin{equation}\eqn{w0_definition}
\left[ t \frac{d}{dt} t^2 \langle E(t) \rangle \right]_{t=w_0^2} = c \ ,
\end{equation}

\noindent
where $w_0$ is the improved scale. 
Again, the value of the fiducial point $c=0.3$ or $c=2/3$ is chosen to avoid discretization and finite-volume effects.

\subsubsection{Chiral-perturbation theory \label{chipt}}
Because both scales $t_0$ and $w_0$ are defined in terms of the energy density $\langle E(t) \rangle$, and the energy density is a local, gauge-invariant quantity, chiral-perturbation theory can be applied to determine the quark-mass dependence of the scales. 
This is an advantage over some other scales, such as $r_0$ or $r_1$, for which no chiral-perturbation theory expansion is available. 
The mapping of $\langle E(t) \rangle$ to the chiral effective theory has been carried out by B\"ar and Golterman in \rcite{chipt}. 
The expansion for $\sqrt{t_0}$ in the $N_f=2+1$ case in terms of the pion and kaon mass is
\begin{eqnarray}
	\sqrt{t_0} & = & \sqrt{t_{0,ch}}\left[ 1 + k_1 \frac{2M_K^2 + M_\pi^2}{(4\pi f)^2} \right. \nonumber\\
		& + & \frac{1}{(4\pi f)^2} \left( (3k_2-k_1)M_\pi^2\mu_\pi + 4k_2M_K^2\mu_K + \frac{1}{3}k_1 (M_\pi^2-4M_K^2)\mu_\eta + k_2M_\eta^2\mu_\eta \right) \eqn{t0-chpt}\\
		& + & \left. k_4\frac{(2M_K^2+M_\pi^2)^2}{(4\pi f)^4} + k_5\frac{(M_K^2-M_\pi^2)^2}{(4\pi f)^4} \right]\ , \nonumber
\end{eqnarray}

\noindent
where $ t_{0,ch}$ is the value of $t_0$ in the chiral limit, the chiral logarithms are represented with the shorthand $\mu_Q = (M_Q/4\pi f)^2 \log{(M_Q/\mu)^2}$, and the $k_i$ are low-energy constants (LECs) that depend on the flow time. 
Note that chiral logarithms enter only at next-to-next-to-leading order (NNLO).  
The scale $w_0$ has the same expansion form to NNLO, but with different coefficients $k_i$. 
This is because the flow-time dependence of $\langle E(t) \rangle$ appears only in the LECs, allowing the differences between \eqs{t0_definition}{w0_definition} to be absorbed into redefinitions of the LECs.

One can generalize \eq{t0-chpt} to staggered chiral-perturbation theory in order to explicitly take into account discretization effects from staggered taste-symmetry violations.  
In this paper, however, we have used simple polynomial expansions to parametrize lattice-spacing effects.  
There are two reasons for this choice. 
First, the quark-mass dependence of the gradient-flow scales is already small, as will be evident in Sec.~\ref{mass}, and nontrivial staggered effects would come in only with the chiral logarithms, which are of NNLO.  
For HISQ quarks, such effects are very small. 
Second, the number of undetermined coefficients in staggered chiral-perturbation theory expansions would be too large in comparison to the number of independent data points available for interpolations. 
Unlike analyses of pseudoscalar masses or decay constants, here we have no valence quarks whose masses could be varied to increase the size of the data set.

\subsubsection{Discretization effects \label{disc}}
In determining the scales $t_0$ and $w_0$, lattice artifacts enter in three places: the action used to generate the initial configurations, the action of the gradient flow, and the choice of observable. 
Because  ensemble generation is expensive, the  action chosen for generating the gauge configurations is fixed in practice. 
Therefore, we only consider improvements to the gradient flow and energy density.

Empirical results suggest partial improvements of the flow or the energy density can yield smaller $O(a^2)$ terms. 
By using the tree-level improved Symanzik action instead of the Wilson action in the flow, the BMW Collaboration found smaller cutoff effects for both gradient-flow scales on their Wilson-clover ensembles with 2-HEX smearing (with scale set by $M_\Omega$) \cite{bmw}. 
Similarly, using the symmetric, cloverleaf definition of the field-strength tensor $G_{\mu\nu}$ in $\langle E \rangle = G_{\mu\nu}G_{\mu\nu}/4$, instead of the simpler sum over the plaquettes, yielded cutoff effects in $\sqrt{t_0}/r_0$ that were five times smaller \cite{t0}. 
Of course,  applying partial improvements at different steps is not guaranteed to produce smaller cutoff effects in the final result. 
Also, for each case, the lattice-spacing dependence of the gradient-flow scale cannot be cleanly separated in the numerical results from the dependence of the additional quantity used to set the scale in the extrapolation to the continuum. 

A detailed examination of the discretization effects on gradient-flow scales has been recently carried out in \rcite{treelevel}. 
The net lattice-spacing dependence from all three stages of the calculation (dynamical action, flow, and observable) is determined at tree level in the gauge coupling from a calculation of $\langle E(t) \rangle$ at finite lattice spacing. 
For the clover observable chosen in this study
 \begin{eqnarray}
 F(t) \equiv\langle t^2 E(t) \rangle &=& \frac{3(N^2-1)g_0^2}{128\pi^2}\left(1+C_2a^2/t + O(a^4/t^2) + O(g_0^2)\right)\ ,\eqn{a2-flow} \\
 C_2 &=& 2c_f+\frac{2}{3}c_g-\frac{1}{24} \ ,\eqn{C2}
\end{eqnarray}

\noindent
where the coefficient $c_f$ describes the gradient-flow action, and $c_g$ describes the original gauge action used to generate the ensembles \cite{treelevel}.  
For our choices of Symanzik one-loop-improved gauge action ($c_g=-1/12$ at tree level) and Symanzik tree-level gradient flow ($c_f=-1/12$), we have $C_2 = -19/72$. 
Unfortunately, our choices of actions and observable lead to larger tree-level discretization terms than from many other combinations of common choices of action for the flow and observable. 
For more details see Table~1 in \rcite{treelevel}.

Utilizing the known $a^2$ dependence of $F(t)$, ``improved" scales are defined in \rcite{treelevel} by canceling the tree-level contributions to $F(t)$ in the implicit definitions of $t_0$ and $w_0$:
\begin{eqnarray}\eqn{improved_t0_definition}
\left[                \frac{t^2 \langle E(t) \rangle}{\left(1 + C_2(a^2/t) + C_4(a^2/t)^2 + ... \right)} \right]_{t=t_{0,{\rm imp}}\phantom{t}}&& = c\ . \\
\left[ t \frac{d}{dt} \frac{t^2 \langle E(t) \rangle}{\left(1 + C_2(a^2/t) + C_4(a^2/t)^2 + ... \right)} \right]_{t=w_{0,{\rm imp}}^2}&& = c \eqn{improved_w0_definition}
\end{eqnarray}

\noindent
For clarity, we will use $t_{0,{\rm orig}}$ and $w_{0, {\rm orig}}$ from here on to refer to the original definition of $t_0$ and $w_0$, and reserve the notations $t_0$ and $w_0$ to refer generically to both the original and improved versions, or to discuss their continuum limits, which is of course common to both versions.
The tree-level improvement in \eqs{improved_t0_definition}{improved_w0_definition} is not obviously an improvement in the nonperturbative region of flow-time where the scales are determined. 
However, the tree-level improvement may be worthwhile if discretization errors arise predominantly from the small-$t$ region, as observed by BMW \cite{bmw}.
We compute the improved scales $t_{0, {\rm imp}}$ and $w_{0, {\rm imp}}$ and compare it to the $a^2$ dependence of the original scales in Sec.~\ref{data}. 

An additional theoretical handle on the comparison can be made by expanding the original scales directly as a power series in $a^2$ and calculating the coefficients.
The lattice-spacing dependence of the gradient-flow scales are proportional to $C_2$ and depend on the continuum flow-time dependence of $F(t)$ and its derivatives $F'(t) = t \frac{d}{dt} F(t)$ and $F''(t) = t^2 \frac{d^2}{dt^2} F(t)$ evaluated at the corresponding continuum scale $t=t_{0}$ or $t=w_{0}^2$. 
The next-to-leading-order coefficients are given by 
\begin{eqnarray}\eqn{tree_cont}
t_{0,{\rm orig}} = t_{0} \left( 1 - T_2 \frac{a^2}{t_{0}} \right), &\hspace{10mm}&T_2 = C_2 \frac{F}{F'} \approx -0.3568(2)\ , \\
w_{0,{\rm orig}}^2 = w_{0}^2 \left( 1 - W_2 \frac{a^2}{w_{0}^2} \right), &\hspace{10mm}&W_2 = C_2 \frac{F'-F}{F''+F'} \approx 0.070(2)\ ,
\eqn{tree_cont2}
\end{eqnarray}

\noindent
Note that the coefficients $T_2$ and $W_2$ are identical to those derived for the improved scale in \rcite{treelevel}; however, the $a^2$ coefficients in the above expression are $-T_2$ and $-W_2$ because \eq{tree_cont} relates the (unimproved) scales at finite lattice spacing to the continuum scales. 
The numerical evaluation of $F$, $F'$, and $F''$ for the estimates of $T_2$ and $W_2$ has been performed on the $a\approx 0.06$ fm, physical quark-mass ensemble (see \tabref{hisqPhys}). 
No systematic errors are included in these estimates.
Unfortunately, because $\sqrt{t_{0,{\rm orig}}}$ and $w_{0,{\rm orig}}$ are defined at flow times outside the perturbative regime, the systematic error on $T_2$ and $W_2$ from higher order and nonperturbative contributions to our estimates of $F$, $F'$, and $F''$ is not known. 
Since discretization errors for $t_{0,{\rm orig}}$ appear to enter primarily at short flow times \cite{bmw}, nonperturbative contributions to $T_2$ may well be small. 
However, there is no corresponding evidence to support a similar conclusion for $W_2$. 
This is discussed further in Sec.~\ref{regress}.

\section{Details of the Computation \label{hisq}}
We compute the scales $\sqrt{t_{0,{\rm orig}}}/a$, $w_{0,{\rm orig}}/a$, $\sqrt{t_{0,{\rm imp}}}/a$, and $w_{0,{\rm imp}}/a$ on the MILC $N_f=2+1+1$ HISQ ensembles \cite{hisq_ensembles, hisq_action}. 
Before describing the gradient-flow simulation details, we tabulate the properties of the ensembles and those quantities needed from prior analyses.
\tabrefs{hisqPhys}{hisqUnphys} list the parameters and relevant observables for ensembles with the strange sea-quark mass tuned near its physical value, and well below its physical value, respectively. 
\tabref{fp4s} gives the values of $aF_{p4s}$ at physical quark masses and associated lattice spacings, which are needed for continuum extrapolations. 
The lattice spacings are calculated with a mass-independent scale-setting scheme; the continuum value $F_{p4s}=153.90(9)({}^{+21}_{-28})$MeV is taken from \rcite{fp4s}, where $f_\pi$ was used to set the absolute scale. 
Physical values of $am_s$ at each lattice spacing \cite{fp4s} are also tabulated. 
Using the physical quark-mass ratio $m_c/m_s = 11.747(19)({}^{+59}_{-43})$ \cite{fp4s}, these values of $am_s$ determine values of the physical charm-quark mass for each ensemble in lattice units, which in turn will be used to adjust for mistunings of the charm sea-quark mass in Sec.~\ref{charm}. 
Finally, \tabref{fp4s} lists the effective coupling constant $\alpha_s$ calculated from taste violations of the HISQ pions in \rcite{fp4s}.
The couplings are scaled by a constant so that $\alpha_s=\alpha_V(q^*=1.5/a)$ for $\beta=5.8$, where $\alpha_V$ is determined from the plaquette \cite{alpha_s,hisq_ensembles}. The values of $\alpha_s$ are used  below in continuum extrapolations.

\begin{table}
\caption{\label{tab:hisqPhys} 
HISQ ensembles with near-physical strange sea-quark mass. 
The first three columns list the gauge coupling constant, the approximate lattice spacing, and the ratio of light- to strange-sea-quark mass. The fourth and fifth columns list the strange and charm sea-quark masses, respectively. 
(Quark masses with primes indicate simulation values of the ensemble, whereas unprimed masses indicate physical values.) 
All but two ensembles can be uniquely identified by the second and third columns. 
To differentiate between the two $a\approx0.12$ fm, $m'_l/m'_s=1/10$ ensembles we use the dimensions of the lattice, $N_s^3 \times N_t$, given in column 6. 
The last two columns give the taste-Goldstone pion and kaon masses in lattice units.}
\begin{tabular}{cccccccc}
\hline\hline
$\beta$ & \ $\approx a$(fm)\  & \ $m_l'/m_s'$\  & $am_s'$ & $am_c'$ & $N_s^3 \times N_t$ & $aM_\pi$ & $aM_K$ \\
\hline
\ $5.80$\  & $0.15$ & $1/5$  & \ $0.0650$ \ & \ $0.838$\  & $16^3 \times 48$  & \ $0.23653(22)$\  & \ $0.40261(25)$\  \\
$5.80$ & $0.15$ & $1/10$ & $0.0640$ & $0.828$ & $24^3 \times 48$  & $0.16614(10)$ & $0.38067(16)$ \\
$5.80$ & $0.15$ & $1/27$ & $0.0647$ & $0.831$ & $32^3 \times 48$  & $0.10180(09)$ & $0.37093(16)$ \\
\hline
$6.00$ & $0.12$ & $1/5$  & $0.0509$ & $0.635$ & $24^3 \times 64$  & $0.18917(15)$ & $0.32358(20)$ \\
$6.00$ & $0.12$ & $1/10$ & $0.0507$ & $0.628$ & $32^3 \times 64$  & $0.13424(09)$ & $0.30813(15)$ \\
$6.00$ & $0.12$ & $1/10$ & $0.0507$ & $0.628$ & $40^3 \times 64$  & $0.13400(06)$ & $0.30821(09)$ \\
$6.00$ & $0.12$ & $1/27$ & $0.0507$ & $0.628$ & $48^3 \times 64$  & $0.08153(04)$ & $0.29851(11)$ \\
\hline
$6.30$ & $0.09$ & $1/5$  & $0.0370$ & $0.440$ & $32^3 \times 96$  & $0.14055(17)$ & $0.24061(18)$ \\
$6.30$ & $0.09$ & $1/10$ & $0.0363$ & $0.430$ & $48^3 \times 96$  & $0.09852(08)$ & $0.22688(12)$ \\
$6.30$ & $0.09$ & $1/27$ & $0.0363$ & $0.432$ & $64^3 \times 96$  & $0.57215(04)$ & $0.21946(09)$ \\
\hline
$6.72$ & $0.06$ & $1/5$  & $0.0240$ & $0.286$ & $48^3 \times 144$ & $0.09438(16)$ & $0.16191(16)$ \\
$6.72$ & $0.06$ & $1/10$ & $0.0240$ & $0.286$ & $64^3 \times 144$ & $0.06713(06)$ & $0.15452(09)$ \\
$6.72$ & $0.06$ & $1/27$ & $0.0220$ & $0.260$ & \ $96^3 \times 192$\  & $0.03887(03)$ & $0.14269(06)$ \\
\hline\hline
\end{tabular}
\end{table}

\begin{table}
\caption{\label{tab:hisqUnphys}
HISQ ensembles with a lighter-than-physical strange sea-quark mass. 
All ensembles have gauge coupling constant $\beta=6.00$ and lattice spacing $a\approx0.12$ fm. 
The first two columns list the approximate values of the light sea-quark mass $m_l'$ and strange sea-quark mass $m_s'$ in units of the physical strange-quark mass $m_s$. 
All of the ensembles may be uniquely identified by these two columns. 
The remaining columns are equivalent to those in \tabref{hisqPhys}. }
\begin{tabular}{cccccc}
\hline\hline
$\approx m_l'/m_s$ & $\approx m_s'/m_s$ & $am_c'$ & $N_s^3 \times N_t$ & $aM_\pi$ & $aM_K$ \\
\hline
$0.10$  & $0.10$ & $\ 0.628\ $ & $32^3 \times 64$ &\  $0.13181(10)\ $ &\   $0.13181(10)$\  \\
$0.10$  & $0.25$ & $0.628$ & $32^3 \times 64$ & $0.13250(09)$ & $0.17385(11)$ \\
$0.10$	& $0.45$ & $0.628$ & $32^3 \times 64$ & $0.13275(10)$ & $0.21719(12)$ \\
$0.10$	& $0.60$ & $0.628$ & $32^3 \times 64$ & $0.13324(10)$ & $0.24509(13)$ \\
\hline
$0.175$ & $0.45$ & $0.628$ & $32^3 \times 64$ & $0.17491(10)$ & $0.23199(12)$ \\
$0.20$  & $0.60$ & $0.635$ & $24^3 \times 64$ & $0.18850(17)$ & $0.26382(18)$ \\
$0.25$  & $0.25$ & $0.640$ & $24^3 \times 64$ & $0.20903(19)$ & $0.20903(19)$ \\
\hline\hline
\end{tabular}
\end{table}

\begin{table}
\caption{\label{tab:fp4s}
Values of $am_s$, $aF_{p4s}$, $a$ (in femtometers), and $\alpha_s$ adjusted to physical values of the quark masses, for various couplings $\beta$. 
All results are from the analysis presented in \rcite{fp4s}. 
The first two columns list the gauge coupling and approximate lattice spacing. 
The next two columns list the strange mass and $F_{p4s}$ in lattice units. 
The lattice spacing from $F_{p4s}=153.90(9)({}^{+21}_{-28})$ MeV, in a mass-independent scheme, is listed in the fifth column. 
The final column tabulates the strong coupling constant $\alpha_s$ determined from the taste splittings (see the text). 
For $am_s$ and $a$, the errors are the sum in quadrature of statistical and systematic errors. 
Only statistical errors are shown for $aF_{p4s}$. }
\begin{tabular}{cccccc}
\hline\hline
$\beta$\  & $\approx a$(fm) & $am_s$ & $aF_{p4s}$ & $a$(fm) & $\alpha_s$ \\
\hline
$\ 5.80\ $ &\  $0.15$\  & \ $0.06863({}^{+53}_{-39})$\  &\  $0.119376(71)$\  & \ $0.15305({}^{+57}_{-41})$\  & \ $0.58801$\  \\
$6.00$ & $0.12$ & $0.05304({}^{+41}_{-30})$ & $0.095403(56)$ & $0.12232({}^{+45}_{-33})$ & $0.53796$ \\
$6.30$ & $0.09$ & $0.03631({}^{+29}_{-21})$ & $0.068570(38)$ & $0.08791({}^{+33}_{-24})$ & $0.43356$ \\
$6.72$ & $0.06$ & $0.02182({}^{+17}_{-13})$ & $0.044237(25)$ & $0.05672({}^{+21}_{-16})$ & $0.29985$ \\
\hline\hline
\end{tabular}
\end{table}

\subsection{Computational setup \label{setup}}
We solve the gradient-flow differential equation numerically using the Runga-Kutta algorithm generalized to SU(3) matrices, as originally proposed by L\"uscher \cite{t0}. 
The routine discretizes the flow time with a step size $\epsilon$ and computes the gauge configuration at a later flow time $t=n\epsilon$ by iterating from the initial gauge configuration. 
The total error of the integration up to flow time $t$ scales like $\epsilon^3$. 
For all ensembles analyzed in this paper, we find that the scales $\sqrt{t_{0,{\rm orig}}}/a$ and $w_{0,{\rm orig}}/a$ determined at a step size of $\epsilon=0.07$ cannot be differentiated, within statistical errors, from those at $\epsilon=0.03$. 
We therefore consider $\epsilon=0.03$ to be a conservative step size, and employ it for all results presented below. 

Both the Wilson and Symanzik tree-level actions for the gradient flow have been implemented and are publicly available in the current release of the {\sc milc} code \cite{code}. 
This computation uses the Symanzik tree-level improved action in the gradient flow and the symmetric, cloverleaf definition of the field-strength tensor $G_{\mu\nu}$ in $\langle E \rangle = G_{\mu\nu}G_{\mu\nu}/4$.
 
\subsection{Measurements of gradient-flow scales \label{data}}
\tabrefs{gfPhys}{gfUnphys} show the results for $\sqrt{t_{0,{\rm orig}}}/a$, $w_{0,{\rm orig}}/a$, $\sqrt{t_{0,{\rm imp}}}/a$, and $w_{0,{\rm imp}}/a$ on the HISQ ensembles. 
The scales $\sqrt{t_{0,{\rm imp}}}/a$ and $w_{0,{\rm imp}}/a$ were improved to $O(a^8)$ at tree level using \eqs{improved_t0_definition}{improved_w0_definition} and the coefficients calculated in \rcite{treelevel} for Symanzik-Symanzik-Clover. 
For the ensembles with the smallest lattice volumes, all configurations are included in the computation. 
As the volumes and cost become larger, a fraction of the configurations are run. 
The configurations in each subset are spaced uniformly across the ensembles, with spacings chosen to help reduce autocorrelations. 
The total number of generated configurations, number of configurations in the gradient-flow calculation, and molecular-dynamics time separation between the included configurations are also tabulated for each ensemble in \tabrefs{gfPhys}{gfUnphys}.

\begin{table}
\caption{\label{tab:gfPhys} 
Values of the gradient-flow scales on the physical strange-quark HISQ ensembles listed in \tabref{hisqPhys}. 
The first two columns are the approximate lattice spacing and the ratio of light to strange sea-quark masses, with the lattice dimensions appended as needed to identify each ensemble uniquely. 
The next column shows the ratio of the number of configurations included in the gradient-flow calculation to the number of configurations in the ensemble. 
The fourth column lists the molecular-dynamics time separation $\tau$ between configurations included in the gradient-flow calculation.
Multiple values are listed for cases where independent streams of the same ensemble did not use the same $\tau$. }
\begin{tabular}{cccccccc}
\hline\hline
\ $\approx a$(fm)\  & \ $m_l'/m_s'$\  & $N_{sim}/N_{gen}$ & \ $\tau$\  & $\sqrt{t_{0,{\rm orig}}}/a$ & $w_{0,{\rm orig}}/a$ & $\sqrt{t_{0,{\rm imp}}}/a$ & $w_{0,{\rm imp}}/a$ \\
\hline
$0.15$ & $1/5$  & \ $1020/1020$\  & $5$ &\  $1.1004(05)$\  &\  $1.1221(08)$\  &\  $0.9857(04)$\  & \ $1.1069(10)$\  \\
$0.15$ & $1/10$ & $1000/1000$ & $5$ & $1.1092(03)$ & $1.1381(05)$ & $0.9932(02)$ & $1.1258(06)$  \\
$0.15$ & $1/27$ & $999/1000$  & $5$ & $1.1136(02)$ & $1.1468(04)$ & $0.9969(02)$ & $1.1361(04)$  \\
\hline
$0.12$ & $1/5$            & $1040/1040$ & $5$   & $1.3124(06)$ & $1.3835(10)$ & $1.2003(05)$ & $1.3870(11)$  \\
$0.12$ & $1/10\ (32^3\times64)$ & $999/1000$  & $5$   & $1.3228(04)$ & $1.4047(09)$ & $1.2100(04)$ & $1.4096(09)$  \\
$0.12$ & $1/10\ (40^3\times64)$ & $1000/1028$ & $5$   & $1.3226(03)$ & $1.4041(06)$ & $1.2098(03)$ & $1.4089(06)$  \\
$0.12$ & $1/27$           & $34/999$    & $140$ & $1.3285(05)$ & $1.4168(10)$ & $1.2152(05)$ & $1.4225(11)$  \\
\hline
$0.09$ & $1/5$  & $102/1011$ & \ $50,60$\  & $1.7227(08)$ & $1.8957(15)$ & $1.6280(08)$ & $1.9053(16)$  \\
$0.09$ & $1/10$ & $119/1000$ & $36$    & $1.7376(05)$ & $1.9299(12)$ & $1.6423(05)$ & $1.9406(12)$  \\
$0.09$ & $1/27$ & $67/1031$  & $32,48$ & $1.7435(05)$ & $1.9470(13)$ & $1.6478(05)$ & $1.9583(13)$  \\
\hline
$0.06$ & $1/5$  & $127/1016$ & $48$ & $2.5314(13)$ & $2.8956(33)$ & $2.4618(12)$ & $2.9049(33)$  \\
$0.06$ & $1/10$ & $38/1166$  & $96$ & $2.5510(14)$ & $2.9478(31)$ & $2.4810(14)$ & $2.9582(30)$  \\
$0.06$ & $1/27$ & $49/583$   & $48$ & $2.5833(07)$ & $3.0119(19)$ & $2.5133(07)$ & $3.0223(19)$  \\
\hline\hline
\end{tabular}
\end{table}

\begin{table}
\caption{\label{tab:gfUnphys}
Values of the gradient-flow scales on the HISQ lighter-than-physical strange-quark ensembles listed in \tabref{hisqUnphys}. 
The first two columns are identical to those in \tabref{hisqUnphys} and are used to identify the ensembles. 
The latter six columns are equivalent to those in \tabref{gfPhys}.}
\begin{tabular}{cccccccc}
\hline\hline
\ $ m_l'/m_s$\  & \ $m_s'/m_s$ \ & $N_{sim}/N_{gen}$ & $\tau$ & $\sqrt{t_{0,{\rm orig}}}/a$ & $w_{0,{\rm orig}}/a$ & $\sqrt{t_{0,{\rm imp}}}/a$ & $w_{0,{\rm imp}}/a$ \\
\hline
$0.10$  & $0.10$ & \ $102/1020$\  & \ $20$\  & \ $1.3596(06)$\  & \ $1.4833(13)$\  & \ $1.2441(06)$\  & \ $1.4932(13)$\   \\
$0.10$  & $0.25$ & $204/1020$ & $20$ & $1.3528(04)$ & $1.4676(10)$ & $1.2378(04)$ & $1.4764(10)$  \\
$0.10$	& $0.45$ & $205/1020$ & $20$ & $1.3438(05)$ & $1.4470(10)$ & $1.2296(05)$ & $1.4544(11)$  \\
$0.10$	& $0.60$ & $107/1020$ & $20$ & $1.3384(08)$ & $1.4351(16)$ & $1.2247(07)$ & $1.4418(17)$  \\
\hline
$0.175$ & $0.45$ & $133/1020$ & $20$ & $1.3385(05)$ & $1.4349(13)$ & $1.2248(05)$ & $1.4415(14)$  \\
$0.20$  & $0.60$ & $255/1020$ & $20$ & $1.3297(06)$ & $1.4170(12)$ & $1.2166(06)$ & $1.4225(12)$  \\
$0.25$  & $0.25$ & $255/1020$ & $20$ & $1.3374(07)$ & $1.4336(14)$ & $1.2236(06)$ & $1.4402(15)$  \\
\hline\hline
\end{tabular}
\end{table}

The error shown with each scale is statistical. 
It is determined by performing a jackknife analysis over the included subset of configurations in each ensemble. 
The jackknife bin size is set to be at least twice the integrated autocorrelation length of the energy density, which is determined in Sec.~\ref{auto}.
In many cases the bin size is larger than would be naively estimated by increasing the bin size until the statistical error plateaus, which is further evidence for the conservative nature of our estimates of autocorrelation lengths.

Considering the low cost and the ease of computation, we originally intended to analyze all configurations from the HISQ ensembles. 
However, the desired statistical accuracy is often reached well before an entire ensemble is analyzed, and the cost, although low compared to configuration generation, is significant enough that analyzing all configurations would be an inefficient use of resources at present. 
If higher-precision scales are needed in the future, it would be straightforward to complete the analysis on the full ensembles.

\subsubsection{Comparison of RHMC and RHMD \label{rhmc}}
\begin{figure}
\includegraphics[scale=0.6]{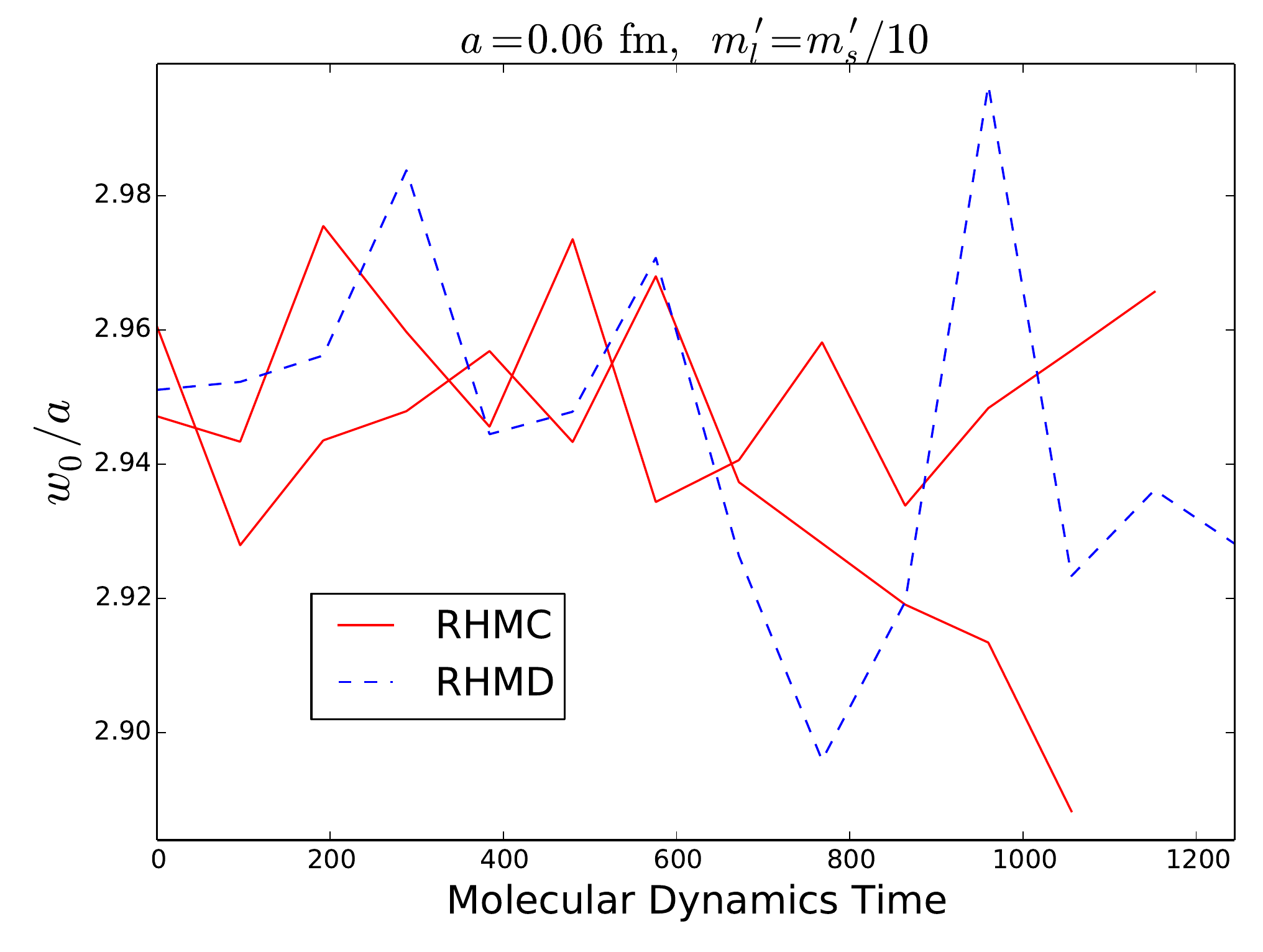}
\includegraphics[scale=0.6]{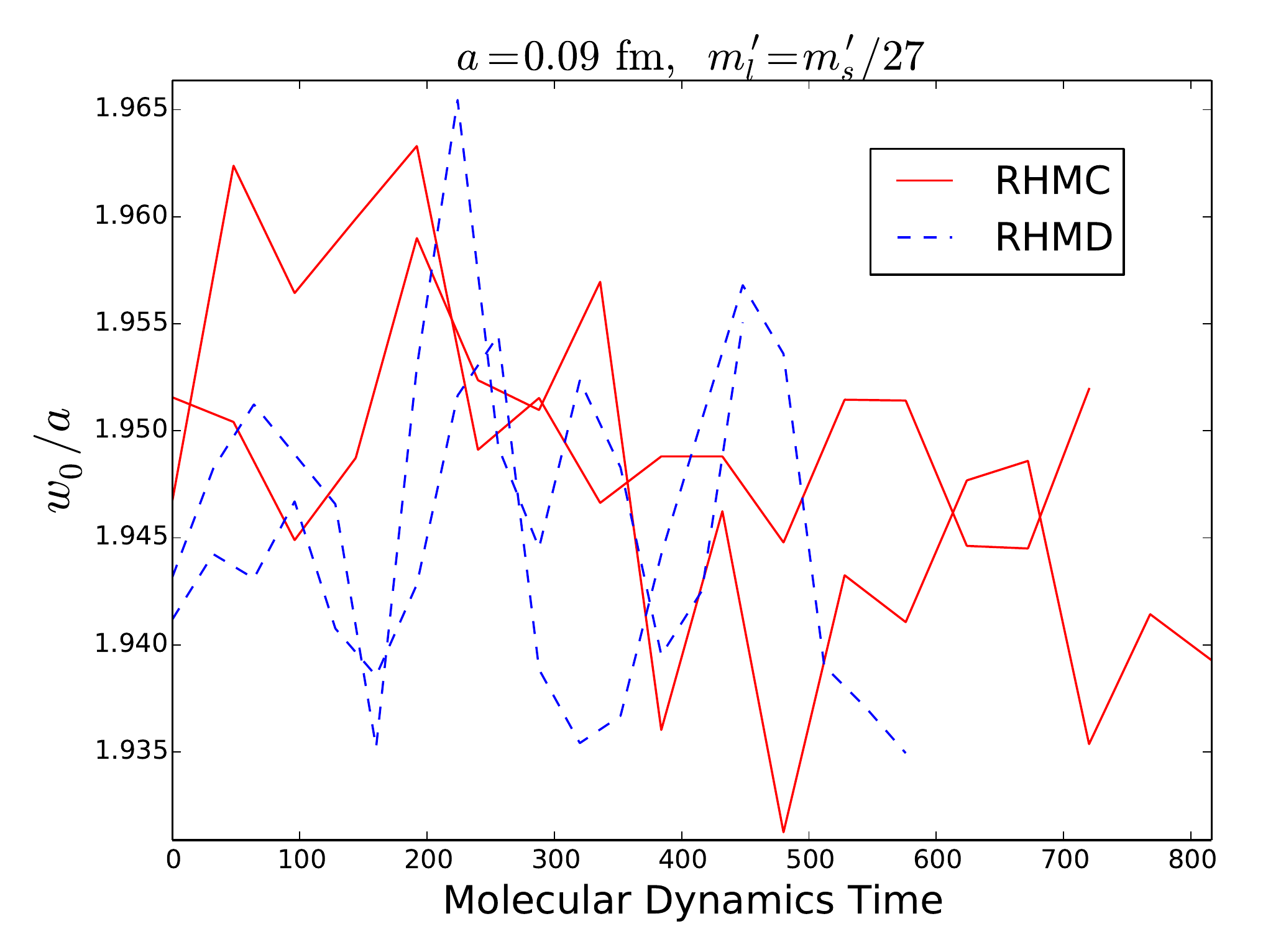}
\caption{\label{rhmcStreams} 
The scale $w_0/a$ measured on individual configurations as a function of the simulation time in molecular-dynamics time units. 
Configuration streams generated with RHMC and RHMD are represented by solid-red and dashed-blue lines, respectively.}
\end{figure}

As discussed in \rcite{hisq_ensembles}, two generation algorithms were employed for the HISQ ensembles: rational hybrid Monte Carlo (RHMC) and molecular dynamics (RHMD). 
As a check of the consistency of these two algorithms, we compute the ratio of $w_0$ computed on RHMC-generated configurations divided by $w_0$ computed on RHMD-generated configurations for the same bare gauge coupling and quark masses. 
For $a \approx 0.09$ fm, $m_l'/m_s' \approx 1/27$, the ratio is $w_0^{RHMC}/w_0^{RHMD}=1.0009(12)$. 
For $a \approx 0.06$ fm, $m_l'/m_s' \approx 1/10$, the ratio is $w_0^{RHMC}/w_0^{RHMD}=1.0002(26)$. 
For some configuration streams the pattern of fluctuations of $w_0/a$ with molecular-dynamics time is not sufficient to reliably estimate the mean and standard deviation over that single stream. 
However, in the particular cases used for calculating the ratio, this issue is not evident. 
Figure \ref{rhmcStreams} shows the fluctuations of the relevant streams for each ratio. 
For all fluctuations of $w_0/a$ on a single stream, the length of the fluctuation in molecular-dynamics time units is small compared to the entire molecular-dynamics time span of the stream.

\subsubsection{Autocorrelation lengths \label{auto}}
The autocorrelation function $\Tau^\Ol(\tau)$ of an observable $\Ol$ is defined as the normalized correlation function of $\Ol$ with itself after the elapse of  molecular-dynamics time $\tau$. 
Given the number of configurations $N$, the ensemble average of the observable $\bar{\Ol}$, and a measurement of the observable $\Ol_i$ on configuration $i$, $\Tau^\Ol(\tau)$ is calculated by
\begin{equation} \eqn{autocorrelation}
\Tau^\Ol(\tau) = \frac{C^\Ol(\tau)}{C^\Ol(0)} \hspace{10mm} C^\Ol(\tau) = \frac{1}{N-\tau} \sum\limits_{i=1}^{N-\tau} \left( \Ol_i - \bar{\Ol} \right) \left( \Ol_{i+\tau} - \bar{\Ol} \right).
\end{equation}
\noindent
The integrated autocorrelation length $l$ is the integral of the autocorrelation function for all cases where $\tau\geq0$. This integral is often estimated as a finite sum using the trapezoidal rule and a cutoff $\tau<\tau_{\rm cut}$, as shown in \rcite{autoSource}:
\begin{equation} \eqn{aclength}
l = \int_0^\infty \Tau^\Ol(\tau)\,d\tau \approx \frac{1}{2} + \sum\limits_{\tau=0}^{\tau_{\rm cut}} \Tau^\Ol(\tau).
\end{equation}
\noindent
The cutoff $\tau_{\rm cut}$ is justified because the autocorrelation function typically decays to 0 as a function of $\tau$ while the statistical noise increases with $\tau$.

To compute statistical errors in the autocorrelation function and integrated autocorrelation length, two independent methods are employed: jackknife and the approximations outlined by Madras and Sokal in \rcite{autoSource}.
For the jackknife method, the ensemble's configurations are divided into distinct bins of $b$ adjacent configurations, and the $j$th jackknife subensemble $\Ol_{\{j\}}$ is defined to be the set of all configurations not contained in bin $j$.
The autocorrelation function $\Tau^\Ol_j(\tau)$ and the integrated autocorrelation length $l_j$ for the $j$th subensemble is computed exactly as for the entire ensemble, except that any contributions involving a configuration from the bin in question are dropped from the sum, and the factor of $N-\tau$ is decreased accordingly.  
Finally, the sample variance for $\Tau^\Ol$ and $l$ is estimated by measuring the variance over the set of jackknife subensembles.
Defining the number of jackknife subensembles to be $M=N/b$, the variance in any quantity $x$ that can be calculated on individual configurations is, from standard jackknife analysis,
\begin{equation}\eqn{jack}
\sigma^2_x = \frac{M-1}{M} \sum\limits_{j=1}^{M} (x_j-\bar{x})^2.
\end{equation}
In applying \eq{jack} to the autocorrelation function, which cannot be estimated from an \emph{individual} configuration, we neglect complications from pairs of configurations between different bins. 
This leads to corrections to \eq{jack} of $O(\tau/b)$, which can be neglected if the bin size is chosen to be large enough that $b\gg\tau_{\rm cut}$.
Intuitively this makes sense because, for sufficiently large sample and bin sizes, the jackknife calculation is similar to breaking one large experiment into several smaller, mostly independent experiments. 
As long as the autocorrelation function can still be computed over its entire domain within any of the smaller experiments, the analogy still holds.

As an additional check that the standard jackknife formulas apply, we used the Metropolis-Hastings algorithm to generate independent streams of Gaussian-distributed real numbers at fixed stream size $N$.
By independently varying the stream size and the number of independent streams within a set, we verified that for a large $N$ and $b>\tau_{\rm cut}$ both the jackknife procedure and the equations of Madras and Sokal yield approximations of the sample variance of $\Tau^\Ol$ and $l$ that agree, within statistical errors, with each other and with the true variance of $\Tau^\Ol$ and $l$.

The second method we employ to estimate the statistical error in $\Tau^\Ol$ and $l$ is the one developed in \rcite{autoSource}.\footnote{Note, that, unlike Madras and Sokal, we call the lag in simulation (molecular-dynamics) time $\tau$, rather that $t$, which is used here for flow time.}
The approximations to the variance neglect $O(l/N)$ effects and, for $\Tau^\Ol$, are slower to compute than the estimates from jackknife. 
However, the jackknife procedure relies on finding a bin size such that $l\ll b\ll N$, which can be tricky for small samples with large correlations.
In the end, we decide to employ and compare both methods because each will introduce different errors as the sample size decreases and the correlation increases.

\begin{figure}
\includegraphics[scale=0.5]{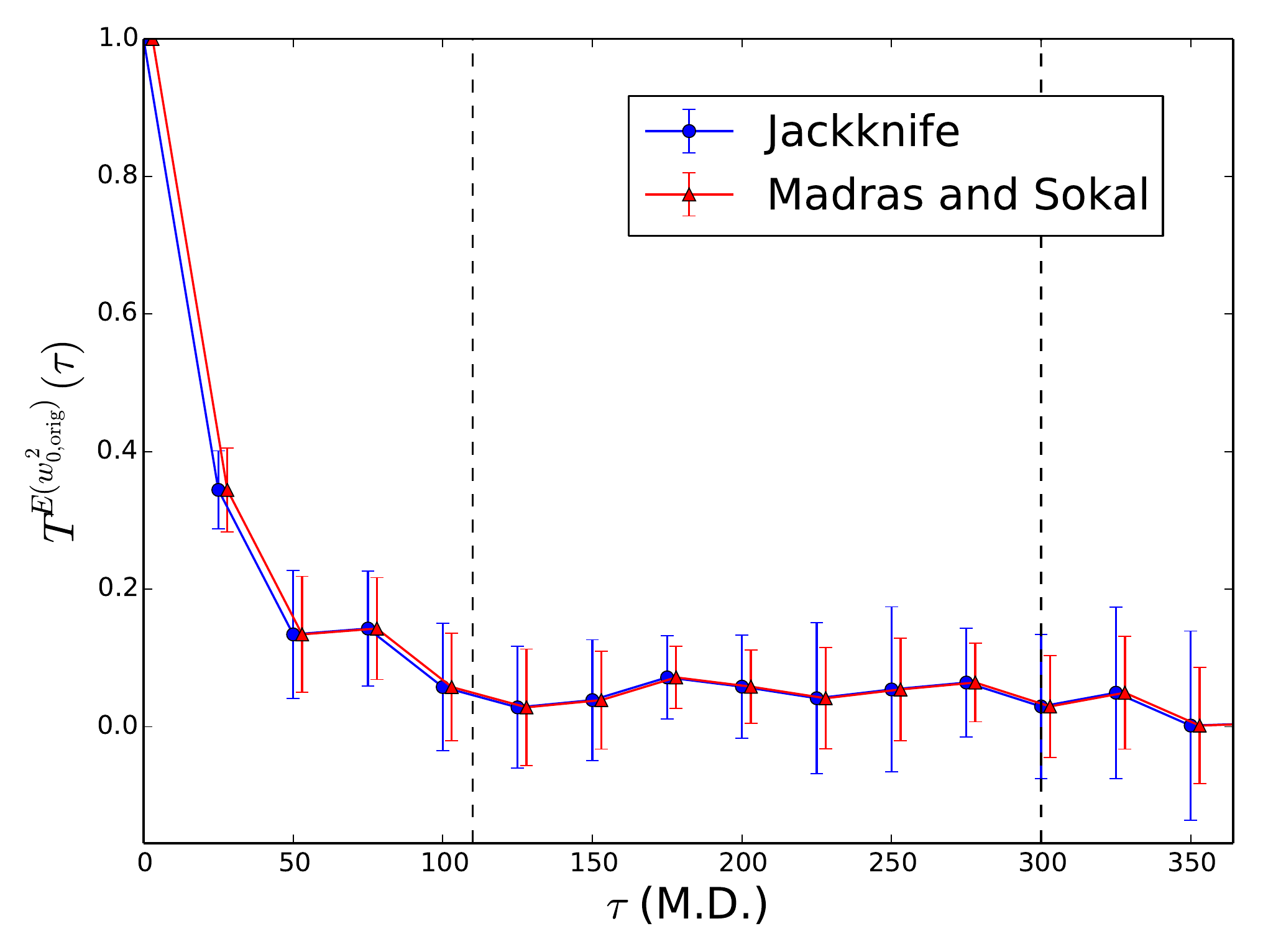}
\includegraphics[scale=0.5]{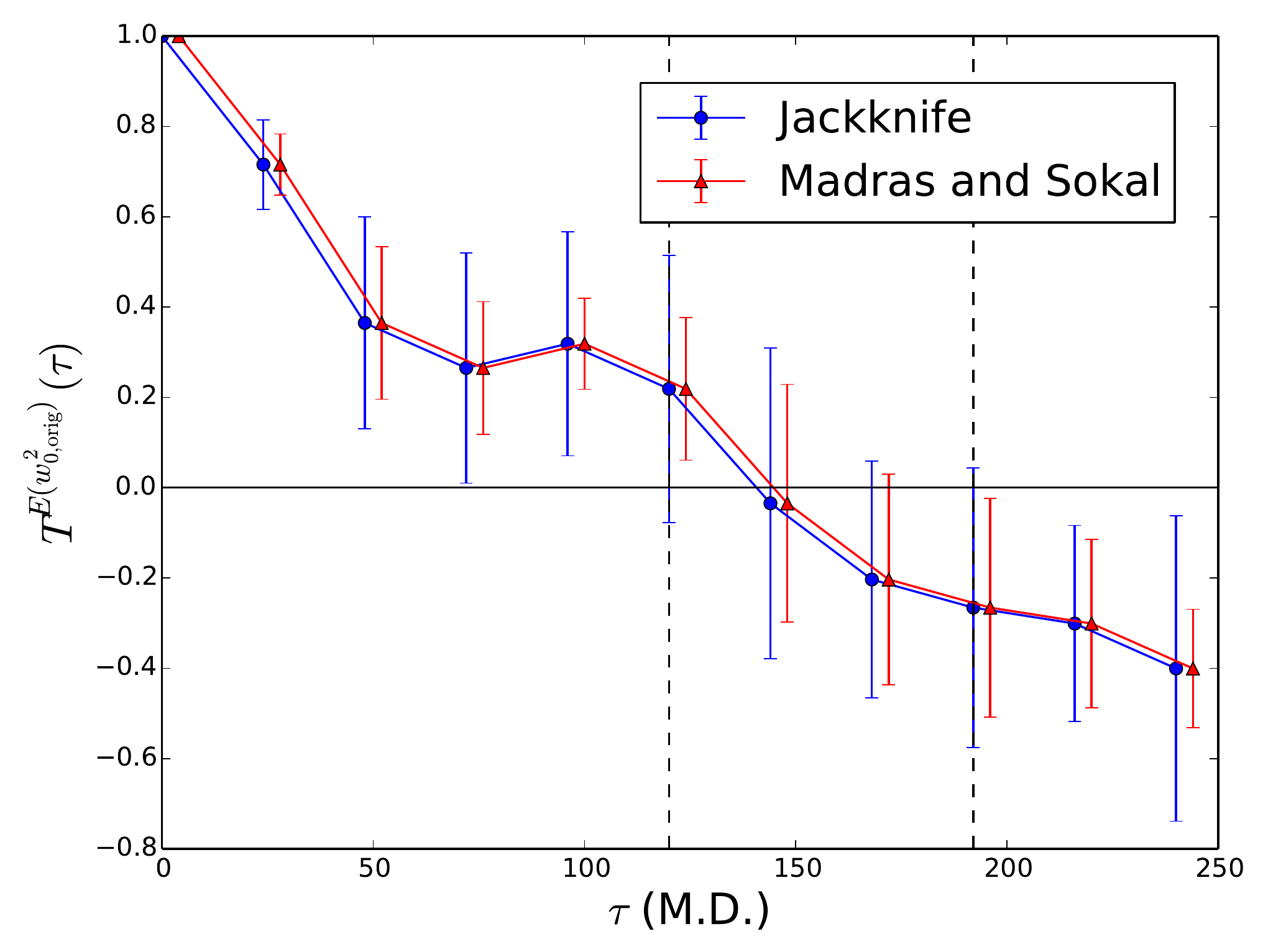}
\caption{\label{fig:acexamples} 
The autocorrelation function of $\langle E(t) \rangle$ plotted as a function of the separation between configurations $\tau$ in molecular-dynamics time units. 
For both figures the flow time $t$ is fixed near the value of $w_0^2$. 
The top plot is for the larger-volume ensemble with $a\approx 0.12$ fm, $m_l' = m_s'/10$. 
For visibility, only every fifth point in $\tau$ is shown.
The bottom plot is for the $a\approx 0.09$ fm, $m_l'=m_s'/10$ ensemble. 
Estimates of the standard error using jackknife and the formulas in \rcite{autoSource} are present for each data point, with the latter shifted slightly right for visibility.
The lower number of analyzed configurations on the finer ensemble leads to a relatively larger statistical error and step size in $\tau$. 
In both plots the vertical dashed lines correspond to the smallest and largest values of $\tau_{\rm cut}$ considered for each ensemble (see the text).}
\end{figure}

We compute the autocorrelation function of $\langle E(t,\tau) \rangle$ as a function of the flow time $t$ and the number, $\tau$, of molecular-dynamics time units separating configurations. 
\Figref{acexamples} shows examples of the autocorrelation function of $\langle E(t,\tau) \rangle$ at fixed flow time $t\approx w_0^2$ for ensembles at $a\approx 0.12$ and $0.09$ fm.
For the ensembles at $a\approx 0.15$ and 0.12 fm, where the full ensembles have been analyzed, we have a reliable estimate of the statistical error of the autocorrelation function for all values of $\tau$. 
For the finer lattice spacings $a\approx 0.09$, 0.06 fm, estimating the autocorrelation functions for $\tau$ smaller than the separations listed in \tabrefs{gfPhys}{gfUnphys} is impossible without calculating the gradient flow on more configurations. 
To address this, we have analyzed an additional 50 equilibrated configurations separated with $\tau=24$ from the $m_l'/m_s=1/10$, $a=0.09$  ensemble. 
Most of these configurations are not included in the calculation of the gradient-flow scales; we keep the configurations used for computing the scales uniformly spread over each ensemble, with constant separation in $\tau$.
With our limited statistics on the $a\approx 0.06$ fm ensembles, we are unable to get useful information on $l$, and we therefore drop those ensembles from further consideration in this subsection.

Once the autocorrelation function of $\langle E(t) \rangle$ is computed, we integrate the function over the separation $\tau$ for each step in flow time $t$.
The statistical error in $l$ is then estimated either using jackknife or the formulas from Madras and Sokal.
For the coarser $a\approx 0.15$ and 0.12 fm ensembles, where $b$ can be chosen to be bigger than $\tau_{\rm cut}$, we find the two estimates agree well with each other (as implied in the top plot of \figref{acexamples}). 
We choose to use the jackknife estimate, which can be computed more rapidly.
To ensure the bin size used in the jackknife procedure is sufficiently large, we first use a bin size $b$ large enough that the statistical error in $\langle E(t,\tau) \rangle$ is (approximately) unchanged with further increases in bin size.
After determining a value for $l$ and a total error $\sigma_l$, we then repeat the calculation with the smallest bin size that obeys $b\geq 2l$ and evenly divides the sample size.
If the new central value and error estimate leads to values of $l$ that do not satisfy this condition, then the bin size is further increased, and this procedure is repeated until the condition is met.
For the finer $a\approx 0.09$ ensemble, a bin size cannot be chosen that falls well between $\tau_{\rm cut}$ and the sample size. So, we choose to use the method of \rcite{autoSource} which yields slightly larger errors (by about $20\%$).

After calculating the statistical error in $l$, the bias from introducing $\tau_{\rm cut}$ must also be accounted for. 
Here we use a slight elaboration on the automatic windowing algorithm mentioned in \rcite{autoSource}: $\tau_{\rm cut}$ is selected to be the lowest value possible that satisfies $\tau_{\rm cut} \geq c l(\tau_{\rm cut})$ for an appropriate choice of $c$. 
Once $c$ is chosen, the remaining bias is approximately equal to $\Tau^\Ol(\tau_{\rm cut}) l(\tau_{\rm cut})$. 
In \rcite{autoSource} a value of $c=10$ was empirically found to yield an acceptable balance of statistical noise and bias; however, our samples are significantly noisier so a smaller value of $c$ is appropriate.
With this in mind, we use the following strategy to come up with our final choice of $\tau_{\rm cut}$.
First, we identify the smallest value of $c=c_{\rm min}$ where $\Tau^\Ol(\tau_{\rm cut})$ is consistent with zero within statistical error. 
We then choose the value of $c$ within the range $c_{\rm min}\leq c \leq 2c_{\rm min}$ that yields the highest $l$.
For the $a\approx 0.15$ and 0.12 fm ensembles, we find $c_{\rm min}=4$ and $c=8$, because the estimates of the autocorrelation functions stay small but positive for a significant range of $\tau$ even after they are first consistent with zero.
For the $a\approx 0.09$ fm ensemble, we find $c_{\rm min}=2$ and $c=2$. This is because the estimate of the autocorrelation function in this case is much noisier and happens to become negative (although consistent with 0) almost immediately after first reaching zero.

\begin{figure}
\includegraphics[scale=0.7]{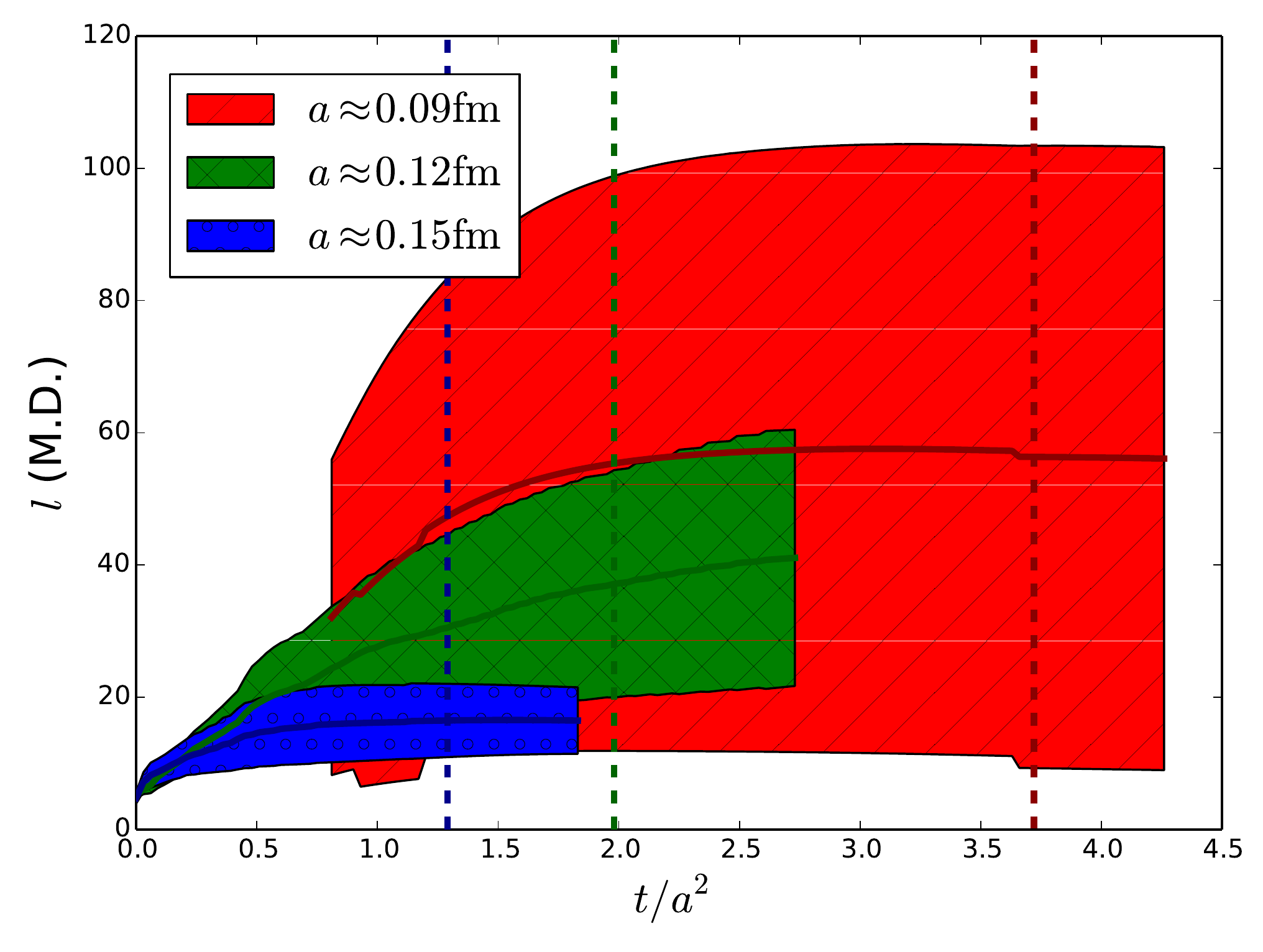}
\caption{\label{fig:avglengths} 
The integrated autocorrelation length (in molecular-dynamics time units) as a function of flow time for ensembles with $m_l'/m_s'=0.1$ and different lattice spacings. 
The thickness of the colored regions show the full range of the $1\sigma$ errors, obtained by adding, in quadrature, the statistical error and bias due to $\tau_{\rm cut}$. 
Dashed vertical lines denote the flow times that determine $w_{0,{\rm orig}}$ on each ensemble where the color of the line matches the color of the shaded region.
For early flow times on the $a\approx 0.09$ fm ensemble the integrated autocorrelation length is not plotted because the smallest step in $\tau$ between configurations is insufficient to resolve the dominant contributions to the autocorrelation function.}
\end{figure}

The integrated autocorrelation lengths with statistical error and the estimated bias combined in quadrature are plotted in \figref{avglengths}.
Notice the autocorrelation length for $\langle E(t) \rangle$ appears to asymptotically increase for increasing flow times, as expected for a smoothing operation. 
The central estimate of the integrated autocorrelation length at large flow times is 58 molecular-dynamics time units for the $a\approx 0.09$ fm, $m_l'=m_s'/10$, physical strange-quark mass ensemble.
In comparison, the integrated autocorrelation length of the topological charge appears to be roughly 40 molecular-dynamics time units for the $a\approx 0.09$ fm, $m_l'=m_s'/5$, physical strange-quark mass ensembles \cite{hisq_ensembles}. 
This suggests the autocorrelation length for $\langle E(t) \rangle$ at large flow times is comparable to the autocorrelation length of the topological charge.

\subsubsection{Charm-quark mass mistuning \label{charm}}

Mistunings of the charm-quark mass on our ensembles vary between  1\%  and 11\%. 
It is therefore important to account for the corrections in the charm-quark mass to the quantities we consider. 
Heavy-quark effects on low-energy quantities come from effects on the scale $\Lambda_{\rm QCD}^{(3)}$ as well as higher-order, physical corrections in powers of $1/m_c$.
Applying only the leading-order corrections from the effect on $\Lambda_{\rm QCD}^{(3)}$ is sufficient for first estimates. 
However, the higher precision of the full continuum extrapolation and quark-mass interpolation requires us to account for the next-to-leading-order (NLO)  contributions, namely the first power corrections in $1/m_c$.
Since the implementation of NLO contributions primarily enters in the full analysis, we defer most of the discussion until Sec.~\ref{regress} and focus here on leading-order effects from the scale $\Lambda_{\rm QCD}^{(3)}$.

If a dimensionless ratio is made of any two quantities evaluated at the same charm-quark mass and with the same dependence on $\Lambda_{\rm QCD}^{(3)}$, then this dependence will cancel in the ratio.  
However, low-energy quantities may also depend on the light-quark masses, which means they may have different dependence on $\Lambda_{\rm QCD}^{(3)}$, even close to the chiral limit.  
Thus, ratios of low-energy quantities may have  leading dependence on $m_c$  from the leftover scale dependence.  
In this analysis we scale all the gradient-flow scales and the meson masses $aM_\pi$, $aM_K$ by the pseudoscalar decay constant $aF_{p4s}$, whose values, adjusted to physical sea-quark masses (including physical $m_c$), are given in \rcite{fp4s}.
Since, for small light-quark masses, $F_{p4s}$ is proportional to $\Lambda_{\rm QCD}^{(3)}$ and the meson masses are proportional to $(\Lambda_{\rm QCD}^{(3)})^{1/2}$, the meson masses must be adjusted to physical $m_c$ to eliminate the leading-order $m_c$ dependence through $\Lambda_{\rm QCD}^{(3)}$.
The gradient-flow scales (in MeV) are also proportional to $\Lambda_{\rm QCD}^{(3)}$ (with quite small sea-quark mass dependence); therefore, scaling by $aF_{p4s}$ evaluated at the same charm-quark mass will cancel the leading-order $m_c$ dependence through $\Lambda_{\rm QCD}^{(3)}$.
To make sure $aF_{p4s}$ and the gradient-flow scales are evaluated at the same charm-quark mass either $aF_{p4s}$ has to be readjusted back to the simulation value $m'_c$ or the gradient-flow scales have to be adjusted to physical $m_c$. 
In this case, we choose to readjust $aF_{p4s}$ to $m'_c$, since its derivative with respect to $m'_c$, $\partial F^2 / \partial m_c = 0.00554(85)$ in {\it p4s} units, has already been computed from the lattice data in \rcite{fp4s}. 
For the ratios of $aM_\pi$ and $aM_K$ to  $aF_{p4s}$,  we  keep the physical-$m_c$ values of $aF_{p4s}$ from  \rcite{fp4s} and adjust $aM_\pi$ and $aM_K$ to the physical value of $m_c$, using the derivative $\partial M^2 / \partial m_c = 0.0209(41)$ calculated in {\it p4s} units \cite{fp4s}.
The values of $aM_\pi$, $aM_K$, and $aF_{p4s}$ after charm-quark-mass adjustments are listed in \tabrefs{charmPhys}{charmUnphys}.

\begin{table}
\caption{\label{tab:charmPhys} 
Results for adjusted meson masses and $aF_{p4s}$, on the physical strange-quark ensembles listed in \tabrefs{hisqPhys}{gfPhys}. 
The adjustments correct for leading-order charm-mass effects, as explained in the text. 
The first two columns are the approximate lattice spacing and ratio of light- to strange-sea-quark mass, with the lattice dimensions appended as needed to uniquely identify each ensemble. 
The next two columns list the masses $aM_\pi$ and $aM_K$ adjusted to the physical charm-quark mass, with the associated statistical error  in parentheses and the change from the data before leading-order charm-quark mass adjustment in square brackets. 
The last column lists the decay constant $aF_{p4s}$ (with  statistical error) adjusted back to the simulation value $am'_c$, while the physical values for the two lighter quark masses are held fixed.}
\begin{tabular}{ccccc}
\hline\hline
\ $\approx a$(fm)\  & $m_l'/m_s'$ & $aM_\pi$ & $aM_K$ & $aF_{p4s}$ \\
\hline
$0.15$ & $1/5$  & \ $0.23619(22)[-34]$\  & \ $0.40204(25)[-57]$\  & \ $0.11976 (7)$\  \\
$0.15$ & $1/10$ & $0.16598(10)[-16]$ & $0.38030(16)[-37]$ & $0.11964 (7)$ \\
$0.15$ & $1/27$ & $0.10169(09)[-11]$ & $0.37051(16)[-41]$ & $0.11967 (7)$ \\
\hline
$0.12$ & $1/5$                       & $0.18904(15)[-13]$ & $0.32335(20)[-22]$ & $0.09555 (6)$ \\  
$0.12$ & \ $1/10\  (32^3\!\times\!64)$\  & $0.13420(09)[-04]$ & $0.30804(15)[-09]$ & $0.09546 (6)$ \\
$0.12$ & $1/10\  (40^3\!\times\!64)$ & $0.13396(06)[-04]$ & $0.30812(09)[-09]$ & $0.09546 (6)$ \\
$0.12$ & $1/27$                      & $0.08151(04)[-02]$ & $0.29843(11)[-08]$ & $0.09546 (6)$ \\
\hline
$0.09$ & $1/5$  & $0.14039(17)[-16]$ & $0.24033(18)[-27]$ & $0.06874 (4)$ \\
$0.09$ & $1/10$ & $0.09849(08)[-03]$ & $0.22681(12)[-07]$ & $0.06861 (4)$ \\
$0.09$ & $1/27$ & $0.05719(04)[-03]$ & $0.21936(09)[-10]$ & $0.06864 (4)$ \\
\hline
$0.06$ & $1/5$  & $0.09400(16)[-38]$ & $0.16125(16)[-65]$ & $0.04465 (3)$ \\
$0.06$ & $1/10$ & $0.06686(06)[-27]$ & $0.15390(09)[-62]$ & $0.04465 (3)$ \\
$0.06$ & $1/27$ & $0.03885(03)[-02]$ & $0.14262(06)[-07]$ & $0.04429 (3)$ \\
\hline\hline
\end{tabular}
\end{table}

\begin{table}
\caption{\label{tab:charmUnphys}
Results for adjusted meson masses and the decay constant $aF_{p4s}$, on the  lighter-than-physical strange-quark ensembles listed in \tabrefs{hisqUnphys}{gfUnphys}.  The adjustments correct for charm-mass mistunings, as explained in the text. The first two columns are identical to those in\tabref{hisqUnphys} and are used to identify the ensembles. The latter three columns are equivalent to those in \tabref{charmPhys}.}
\begin{tabular}{cccccc}
\hline\hline
\ $ m_l'/m_s$\  & \ $m_s'/m_s$ \ & $aM_\pi$ & $aM_K$ & $aF_{p4s}$ \\
\hline
$0.10$  & $0.10$ & \ $0.13177(10)[-04]$ \ & \ $0.13177(10)[-04]$\  & \ $0.09546 (6)$\  \\
$0.10$  & $0.25$ & $0.13247(09)[-04]$ & $0.17380(11)[-05]$ & $0.09546 (6)$ \\
$0.10$	& $0.45$ & $0.13271(10)[-04]$ & $0.21713(12)[-06]$ & $0.09546 (6)$ \\
$0.10$	& $0.60$ & $0.13320(10)[-04]$ & $0.24502(13)[-07]$ & $0.09546 (6)$ \\
\hline
$0.175$ & $0.45$ & $0.17487(10)[-05]$ & $0.23192(12)[-07]$ & $0.09546 (6)$ \\
$0.20$  & $0.60$ & $0.18837(17)[-13]$ & $0.26364(18)[-18]$ & $0.09555 (6)$ \\
$0.25$  & $0.25$ & $0.20883(19)[-21]$ & $0.20883(19)[-21]$ & $0.09561 (6)$ \\
\hline\hline
\end{tabular}
\end{table}

It is instructive to compare these results to the leading-order $m_c$ effect on $\Lambda_{\rm QCD}^{(3)}$ expected from perturbation theory.
Defining the ratio $P(m_c/\Lambda_{\rm QCD}^{(4)}) = \Lambda_{\rm QCD}^{(3)}/\Lambda_{\rm QCD}^{(4)}$, a renormalization-group invariant $\eta(m_c)$ can be constructed from the logarithmic derivative of $P$ \cite{mc_corr}
\begin{equation}\eqn{logP}
\eta(m_c) \equiv \frac{m_c}{\Lambda_{\rm QCD}^{(4)}} \frac{P'}{P},
\end{equation}
with $P'$ being the derivative of $P$ with respect to its argument.
At leading perturbative order, $\eta\approx\eta_0 = 2/27$ \cite{charm_origin, *charm_erratum, charm_text}, and
\begin{equation} \eqn{Lambda3}
\frac{\partial \Lambda_{\rm QCD}^{(3)}}{\partial m_c} = P'\left( \frac{m_c}{\Lambda_{\rm QCD}^{(4)}} \right) \approx \frac{2}{27} \frac{\Lambda_{\rm QCD}^{(3)}}{m_c}.
\end{equation}
Then, given $Q=k(\Lambda_{\rm QCD}^{(3)})^p$, where $k$ and $p$ are independent of $m_c$, the partial derivative of $Q$ with respect to $m_c$ at leading order in perturbation theory (and neglecting physical, NLO corrections in $1/m_c$) is 
\begin{equation} \eqn{dQdmc}
\frac{\partial Q}{\partial m_c} = kp \left( \Lambda_{\rm QCD}^{(3)} \right)^{p-1} \frac{\partial \Lambda_{\rm QCD}^{(3)}}{\partial m_c} \approx \frac{2p}{27} \frac{Q}{m_c}\ .
\end{equation}
As mentioned in \rcite{fp4s}, the results of this formula for $\partial F^2 / \partial m_c$ and $\partial M^2 / \partial m_c$ agree, within 10\%, with the numerical determination of these derivatives from our lattice data.
Also, we find that the dimensionless product of $aF_{p4s}$ with the gradient-flow scales is approximately the same whether $aF_{p4s}$ is readjusted to  $m'_c$ using the numerically estimated derivative or the gradient-flow scales are adjusted to the physical value of $m_c$ using \eq{dQdmc}.
The largest difference between the two approaches is $1.7\sigma_{\rm stat}$, where $\sigma_{\rm stat}$ is the statistical error,  and occurs on the $a\approx 0.06$ fm, $m_l'=m_s'/5$ ensemble for the dimensionless combination $\sqrt{t_{0,{\rm orig}}}F_{p4s}$.

We account for the remaining physical, NLO corrections in powers of $1/m_c$ by directly including such terms in the fits to $\sqrt{t_{0,{\rm orig}}}F_{p4s}$, $w_{0,{\rm orig}}F_{p4s}$, $\sqrt{t_{0, {\rm imp}}}F_{p4s}$, and $w_{0, {\rm imp}}F_{p4s}$. 
Specific details of what powers are included and how the terms are constrained is detailed in Sec.~\ref{regress}. 
The effects of the NLO charm-mass corrections to the meson masses $aM_\pi/aF_{p4s}$ and $aM_K/aF_{p4s}$ on the gradient-flow scales are negligible  because these are quite small corrections and the dependence of the gradient-flow scales on $aM_\pi/aF_{p4s}$ and $aM_K/aF_{p4s}$ is already weak.

\subsubsection{Simple continuum extrapolation \label{naive}}
A simple continuum extrapolation can be quickly performed by including only the physical quark-mass ensembles. 
With just these ensembles, light-quark, strange-quark, and NLO charm-quark-mass mistuning effects cannot be accounted for, and the statistical error will be larger than from a fit to the complete data set. 
Nevertheless, this extrapolation is useful because it provides a  check on the final value from the more complicated fits and highlights the degree of improvement in discretization errors of $w_{0,{\rm orig}}F_{p4s}$ over $\sqrt{t_{0,{\rm orig}}}F_{p4s}$, as well as $\sqrt{t_{0,{\rm imp}}}F_{p4s}$ and $w_{0,{\rm imp}}F_{p4s}$ over the originals $\sqrt{t_{0,{\rm orig}}}F_{p4s}$ and $w_{0,{\rm orig}}F_{p4s}$.

To perform the continuum extrapolation we multiply by the values of $aF_{p4s}$ listed in \tabref{fp4s} to create a dimensionless quantity that is finite in the continuum limit. 
We choose $aF_{p4s}$ to keep the statistical errors smaller than what they would be from an experimentally accessible quantity such as $f_\pi$. 
To convert the final result to physical units, however, we must use  $F_{p4s}=153.90(09)({}^{+21}_{-28})$ MeV, which was computed with the scale set by $af_\pi$. 
The advantage of using $aF_{p4s}$ to set the intermediate scale is that it yields smaller relative scale errors from different ensembles, and thus aids in the extrapolation to the continuum.

\begin{figure}
\includegraphics[scale=0.7]{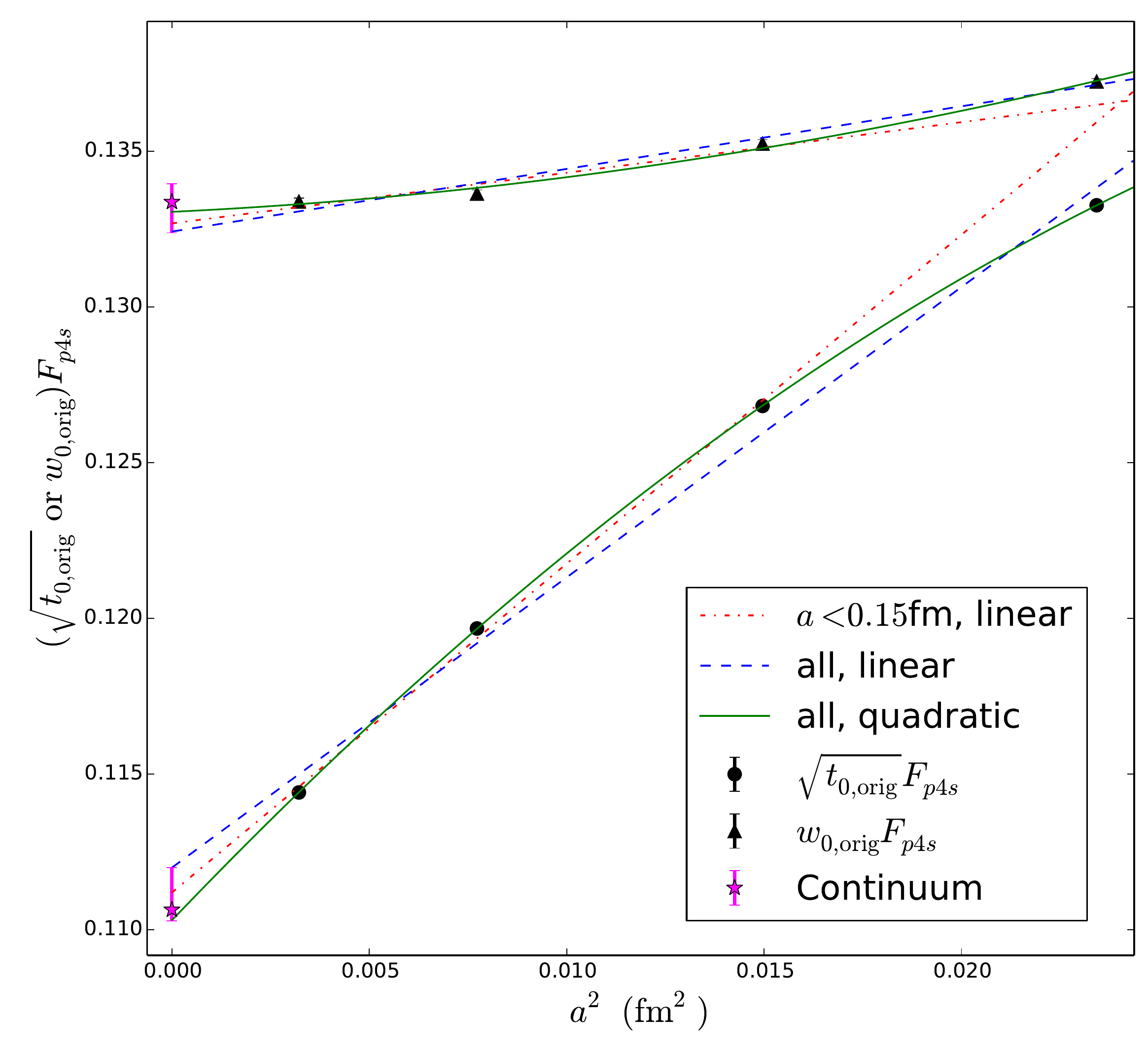}
\caption{\label{fig:naiveTW} 
Simple continuum extrapolations for $\sqrt{t_{0,{\rm orig}}} F_{p4s}$ and $w_{0,{\rm orig}}F_{p4s}$ over  physical quark-mass ensembles only. 
Statistical error bars are present, but they are nearly invisible on this scale. 
Three fits to each data set are shown. 
The red, dot-dashed line is a linear fit in $a^2$ to the three finer ensembles ($a<0.15$ fm), the blue dashed line is a linear fit in $a^2$ to all four ensembles, and the green solid line is a quadratic fit in $a^2$ to all four ensembles. 
The continuum extrapolation points, calculated from $\sqrt{t_{0,{\rm imp}}}$ and $w_{0,{\rm imp}}$, are shown in magenta with error bars representing the sum of statistical and systematic uncertainties in quadrature.}
\end{figure}	

\begin{figure}
\includegraphics[scale=0.7]{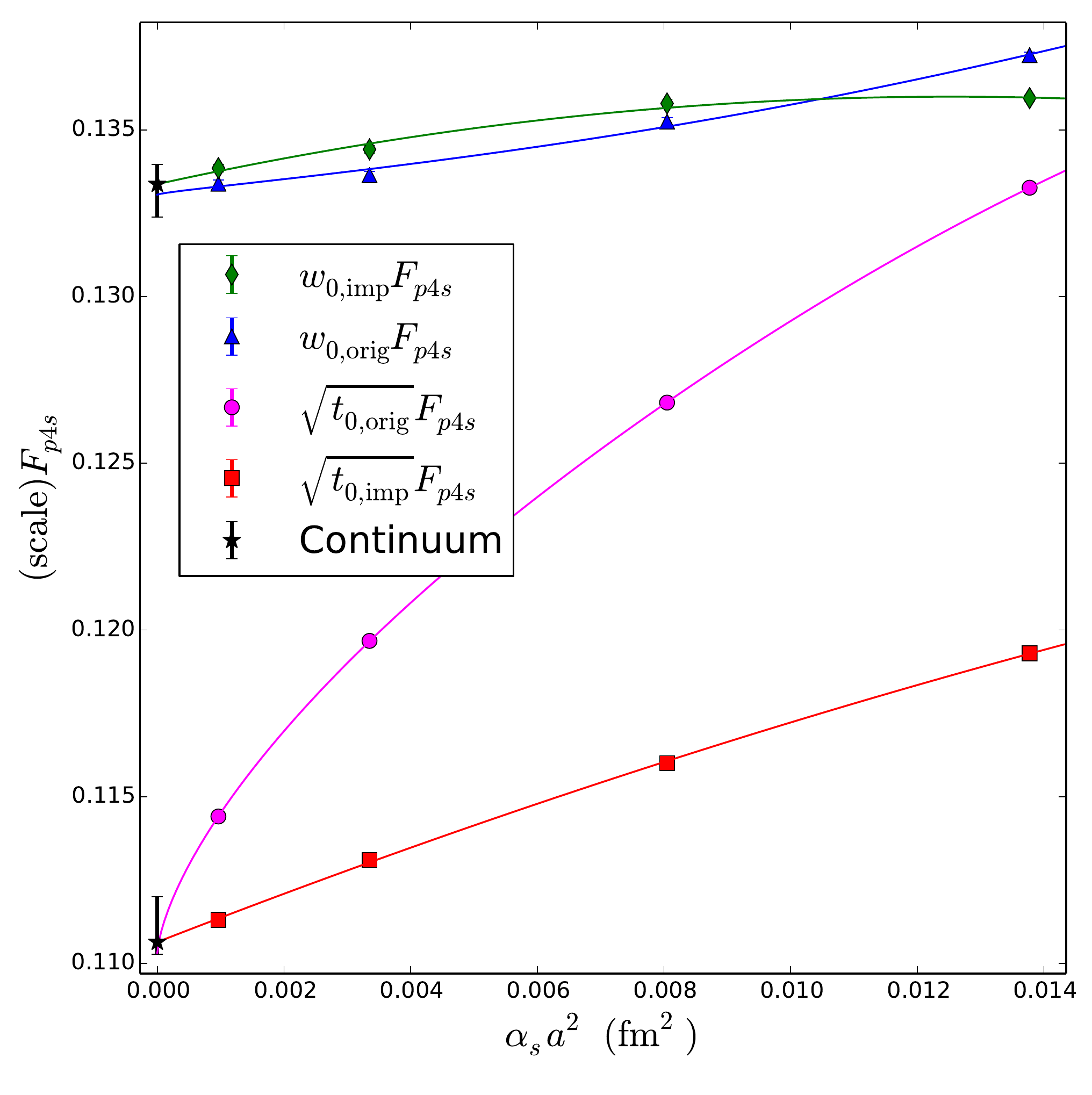}
\caption{\label{fig:naiveImp} 
Simple continuum extrapolations for the original ($\sqrt{t_{0,{\rm orig}}}$ and $w_{0,{\rm orig}}$) and improved ($\sqrt{t_{0,{\rm imp}}}$ and $w_{0,{\rm imp}}$) gradient-flow scale times $F_{p4s}$ over only physical quark-mass ensembles. 
Quadratic fits in $a^2$ or $\alpha_sa^2$ over all four physical-mass ensembles are shown for the original and improved scales, respectively.
The continuum extrapolation points, calculated from the improved scales, are shown in black with error bars representing the sum of statistical and systematic uncertainties in quadrature.}
\end{figure}	

Plots of $\sqrt{t_{0,{\rm orig}}}F_{p4s}$ and $w_{0,{\rm orig}}F_{p4s}$ as a function of $a^2$ are shown in \figref{naiveTW}. 
The discretization improvement of $w_{0,{\rm orig}}$ over $\sqrt{t_{0,{\rm orig}}}$ is immediately evident in the differences between the coarsest and finest ensembles.
This result holds for many  choices of the reference scale, including  $af_\pi$, $r_1/a$, $\sqrt{t_{0,{\rm imp}}}/a$, and $w_{0,{\rm imp}}/a$, in addition to $aF_{p4s}$, the choice used in \figref{naiveTW}.
In addition, the plot shows that the $a^2$ dependence is not trivial for $w_0F_{p4s}$.
This is not unexpected because we are using a highly improved configuration action (which directly affects $aF_{p4s}$) for a statistically  precise measurement. 
The importance of higher-order terms in $a^2$ and $\alpha_s a^2$ can be seen directly in the differences between the improved and original $w_0$, as well as the difference between $\sqrt{t_{0,{\rm orig}}}$ and $w_{0,{\rm orig}}$. 
The situation is further complicated by effects of quark-mass mistunings between ensembles with approximately the same ratio $m_l/m_s$. 
This is explored in more detail in the full fit analysis in Sec.~\ref{fitresults}. 
For now, we include linear fits in $a^2$ with or without the coarsest $a\approx0.15$ fm ensemble and quadratic fits in $a^2$ to all four ensembles.

\Figref{naiveImp} compares the improved scales $\sqrt{t_{0,{\rm imp}}}$ and $w_{0,{\rm imp}}$ with the original ones $\sqrt{t_{0,{\rm orig}}}$ and $w_{0,{\rm orig}}$.
As before, we consider linear fits with or without the coarsest ($a\approx0.15$ fm) ensemble, and quadratic fits with all four ensembles. 
For the unimproved scales the fit curves are functions of $a^2$; for the improved scales they are functions of $\alpha_sa^2$, since tree-level
discretization errors have been removed. 
The improvement at tree level is clear for $\sqrt{t_0}$, where the $\alpha_sa^2$ dependence of $\sqrt{t_{0,{\rm imp}}}$ is close to linear, and the slope is considerably less steep than for $\sqrt{t_{0,{\rm orig}}}$.   
The difference between $w_{0,{\rm orig}}$ and $w_{0,{\rm imp}}$ is much smaller, and is contaminated here by mistuning effects, so we postpone discussion until after we correct for such mistunings.

The continuum values are extracted from the quadratic fit in $\alpha_sa^2$ to the full data set on the improved scales. 
The systematic error from the extrapolation is estimated by the largest differences between this fit and the other fits considered. 
This yields the simple estimates for the gradient-flow scales $\sqrt{t_0} = 0.1419(2)({}_{-\phantom{1}4}^{+17})$ fm and $w_0 = 0.1710(4)({}_{-12}^{+\phantom{1}7})$ fm.
Here we do not include errors (statistical or systematic) from the determination of $F_{p4s}$ so that we can make a cleaner comparison with the extrapolations over the full data set (nonphysical quark masses included) in the next section.

\subsection{Full continuum extrapolation \label{regress}}
Using all of the ensembles listed in \tabrefs{hisqPhys}{hisqUnphys}, we now perform a combined continuum extrapolation and interpolation to physical quark masses. 
Compared with the simple continuum extrapolation over the physical quark-mass ensembles only, the full approach has greater statistics, provides a handle for precise tuning of the light-quark and strange-quark masses to their physical values, and allows for better control and analysis of the systematic errors from discretization effects. 

We break the analysis into two main sections. 
First, the functional forms and parameter variations for controlling mass and lattice-spacing dependence are outlined. 
Second, we present the results from our fits of the lattice data to the models from the first section.

\subsubsection{Models of mass and lattice-spacing dependence\label{fitforms}}
To perform the combined continuum extrapolation/quark-mass interpolation there are three functional forms that must be chosen: quark-mass terms, lattice-spacing terms, and terms that combine both (cross terms).

For the light and strange mass dependence we use the chiral expansion outlined in Sec.~\ref{chipt} with $M_\pi$ and $M_K$ as independent variables, standing in for the light- and strange-quark-mass dependence. 
For each fit we include the expansion up to LO (just a constant), NLO (which adds an analytic 
term linear in the squared meson masses, but no chiral logarithms), or NNLO (chiral logarithms and terms up to quadratic in the squared 
meson masses). 
In the fits to \eq{t0-chpt}, the rho meson mass is used for $\mu$, $f_\pi$ is used for $f$ at NLO, and $F_{p4s}$ is used for $f$ at NNLO for convenience; other choices for these quantities would be equivalent up to redefinitions of $k_i$ and the addition of terms of higher than NNLO order.

For the NLO charm-quark-mass dependence, \rcite{mc_corr} argues that, for large $m_c$, corrections start at $\cO(1/m_c^2)$.  In our central fits we therefore add a term proportional to $1/m_c'^2 - 1/m_c^2$ (primes denote simulation values, unprimed quantities denote physical values).  However, in the lighter-than-charm region where \rcite{mc_corr} performed simulations, their data actually was better described by $1/m_c$ dependence than by $1/m_c^2$.  Although all our values of $m'_c$ are closer to $m_c$ than those of \rcite{mc_corr},  we also consider fits that replace  $1/m_c'^2 - 1/m_c^2$ by $1/m_c'- 1/m_c$ in estimating systematic errors.

For the lattice-spacing dependence we use a Taylor-series ansatz in powers of $a^2$, $\alpha_sa^2$, and  $\alpha_s^2a^2$. 
We include powers of $\alpha_s$ because the leading errors coming from the action in a joint expansion in $a^2$ and $\alpha_s$ is $\alpha_sa^2$ and the leading taste-violating errors is $\alpha_s^2a^2$.
For the original scales $\sqrt{t_{0,{\rm orig}}}$ and $w_{0,{\rm orig}}$, the first order term in lattice spacing, $a^2$, is always included. 
Higher orders are optionally included up to $a^6$, $\alpha_sa^2$, and $\alpha_s^2a^2$. 
For the improved scales $\sqrt{t_{0,{\rm imp}}}$ and $w_{0,{\rm imp}}$,  the first order in either $\alpha_sa^2$ or $a^4$ is always included.
Higher orders are optionally included up to $a^8$, $(\alpha_sa^2)^2$, and $\alpha_s^2a^2$. 
Even though the scales $\sqrt{t_{0,{\rm imp}}}$ and $w_{0,{\rm imp}}$ are improved to order $a^8$ at tree level, the $a^4$ through $a^8$ terms are included for fits because $aF_{p4s}$ has leading corrections of $\alpha_sa^2$ and $a^4$. 
For both scales, the number of lattice-spacing terms in a single fit is not allowed to exceed 3. 
Together with the value of the scale in the continuum limit, this ensures that at most four parameters describe the $a$ dependence of the data from our four unique lattice spacings.

In order to limit the large number of cross terms possible, we only include products of chiral and lattice-spacing terms whose total ``order" is no higher than the largest noncross term included in the fit function. Here by order we simply mean the total power of any of the following factors, which all have similar magnitudes for the HISQ ensembles: 
$\alpha_s \sim (\Lambda_{\rm QCD}a)^2 \sim (M/(4\pi f))^2$.
Also, no cross terms are constructed from the highest orders of mass or lattice-spacing terms. 
For example, a fit including $a^6$ and the chiral expansion to NNLO would include a term like $a^4(M/(4\pi f))^2$ but not $a^2(M/(4\pi f))^4$. 

Once the functional form is chosen, we also consider various restrictions of the data set.
As already suggested from the naive fit to the physical quark-mass ensembles only, the $a\approx0.15$ fm ensembles may require higher orders of $a^2$ to be included. 
So we consider fits that include or drop these ensembles. 
Furthermore, when the $a\approx0.15$ fm ensembles are dropped, we do not include more than two lattice spacing terms to ensure the three unique lattice spacings represented by the data set are parametrized by three or fewer variables.
A second restriction on the data set is determined by the kaon mass. 
The lighter-than-physical strange-quark ensembles have strange-quark masses all the way down to $1/10$ the physical strange-quark mass. 
Including these ensembles along with the physical-mass ensembles that comprise most of out data requires more complex chiral forms to cover the large range in $m'_s$.  
We therefore consider eight different lower bounds for the kaon masses included in the fit, ranging from just below the physical strange-quark mass, to near zero, which includes all the ensembles.
We do not set an upper bound for the kaon mass, as would be typical of chiral-perturbation-theory extrapolations, because this would only leave $a\approx0.12$ fm ensembles for the extrapolation.

We add Gaussian priors centered around zero to ensure the magnitudes of fit parameters are physically plausible; we refer to the standard deviation of the Gaussian as the {\it prior width}.  
For discretization terms of the form $k (\alpha_s^p a^2 \Lambda_{\rm QCD}^2)^n$, the dimensionless coefficient $k$ is presumed to be of order unity so that the finite Tayler-series expansion in $a^2$, $\alpha_s a^2$, and $\alpha_s^2 a^2$ is justified.
When reexpressed in terms of the two dimensionless quantities,
\begin{equation}
\chi_{ud} = \frac{2 \mu m_l'}{8\pi^2f_\pi^2}, \hspace{2cm} \chi_{s} = \frac{2\mu m_s'}{8\pi^2f_\pi^2},
\end{equation}
\noindent
the coefficients of the terms from chiral-perturbation theory are also expected to be of order unity.
\footnote{Prior widths on cross terms are set to the product of the widths associated with each of the factors.}  
A prior width of 1 in these units is in most cases sufficient to ensure that the data, rather than the prior assumption, is constraining a given parameter, since the deviation of the parameter from zero is well within one prior width.  Once the priors widths are increased to 3, this is true for all the discretization and chiral parameters, and most fit results are negligibly different from those with no prior constraints at all. 
The only exceptions are seven fits to $\sqrt{t_{0,{\rm imp}}}F_{p4s}$ whose continuum results differ by $\approx 2\sigma_{\rm stat}$ from those without prior constraints.
 
For the NLO charm-quark-mass corrections, the prior width is based on the results of \rcite{mc_corr}, which finds that such heavy-quark effects on the gradient-flow scales are $\ltwid0.3\%$. 
For a dimensionless ratio $R(m_c)$, we choose the prior width such that $|R(m_c)-R(\infty)|/R(\infty) < 0.5\% $ or $1.5\% $.
Most fits show negligible difference between a prior width of $1.5\%$ and a prior width set to infinity.  
However, the prior width does significantly constrain the $m_c$ dependence on a few outlying fits; without any prior constraints these fits would have shown differences of $4\%$ to $5\%$ between a physical $m_c$ and an infinite $m_c$.  
We consider such a large $m_c$ dependence unreasonable and we believe we are justified in removing these few outliers using the prior constraints.
It is probable that these large NLO $1/m_c$ power corrections are mimicking the dependence on other variables such as $m_s$;
we note that the mistunings in $m_c$ are comparable to and correlated with the mistunings in $m_s$ on the physical strange-quark-mass ensembles.
Another possibility is that the power corrections are making up for errors in the derivatives $\partial F^2 / \partial m_c$ and $\partial M^2 /\partial m_c$.
We have checked to see, however, that varying the derivatives by $2\sigma_{\rm stat}$ does not produce significant variations in the continuum results.
For this reason and others discussed in later sections, we do not widen the prior on NLO charm-quark-mass dependence any further than $1.5\%$ in the final analysis.

This leads us to consider two sets of Gaussian priors in our final analysis: one set with all prior widths set to the smaller choice (1 for discretization and chiral terms, and 0.5\% for NLO charm-mass dependence) and another set with all widths set to the larger choice (3 for discretization and chiral terms, and 1.5\% for NLO charm-mass dependence). Both sets of priors are in general wide enough that the parameters are determined by the data and not the priors (the deviation of the parameter from zero does not change an appreciable fraction of the original width when the width is increased by a factor of 3); the only exceptions are for the parameters determining NLO charm-mass dependence, and then only for a few outlying fits, as described earlier.

For all scales, there are three chiral expansions, eight choices of lower bound for the kaon mass, two choices for the next-to-leading-order charm-quark-mass correction, and two sets of priors. 
For the original scales $\sqrt{t_{0,{\rm orig}}}$ and $w_{0,{\rm orig}}$, there are six lattice-spacing expansions with the  $a\approx0.15$fm ensembles included and three lattice-spacing expansions with the  $a\approx0.15$ fm ensembles {\it not} included. 
This produces a total of $3\times8\times2\times2\times(6+3)=864$ different fits. 
For the improved scales $\sqrt{t_{0,{\rm imp}}}$ and $w_{0,{\rm imp}}$, there are nine lattice-spacing expansions with the $a\approx0.15$fm ensembles included and five lattice-spacing expansions with the $a\approx0.15$ fm ensembles {\it not} included.
This produces a total of $3\times8\times2\times2\times(9+5)=1344$ different fits.

\subsubsection{Fits to the lattice data \label{fitresults}}
\begin{figure}
\includegraphics[scale=0.58]{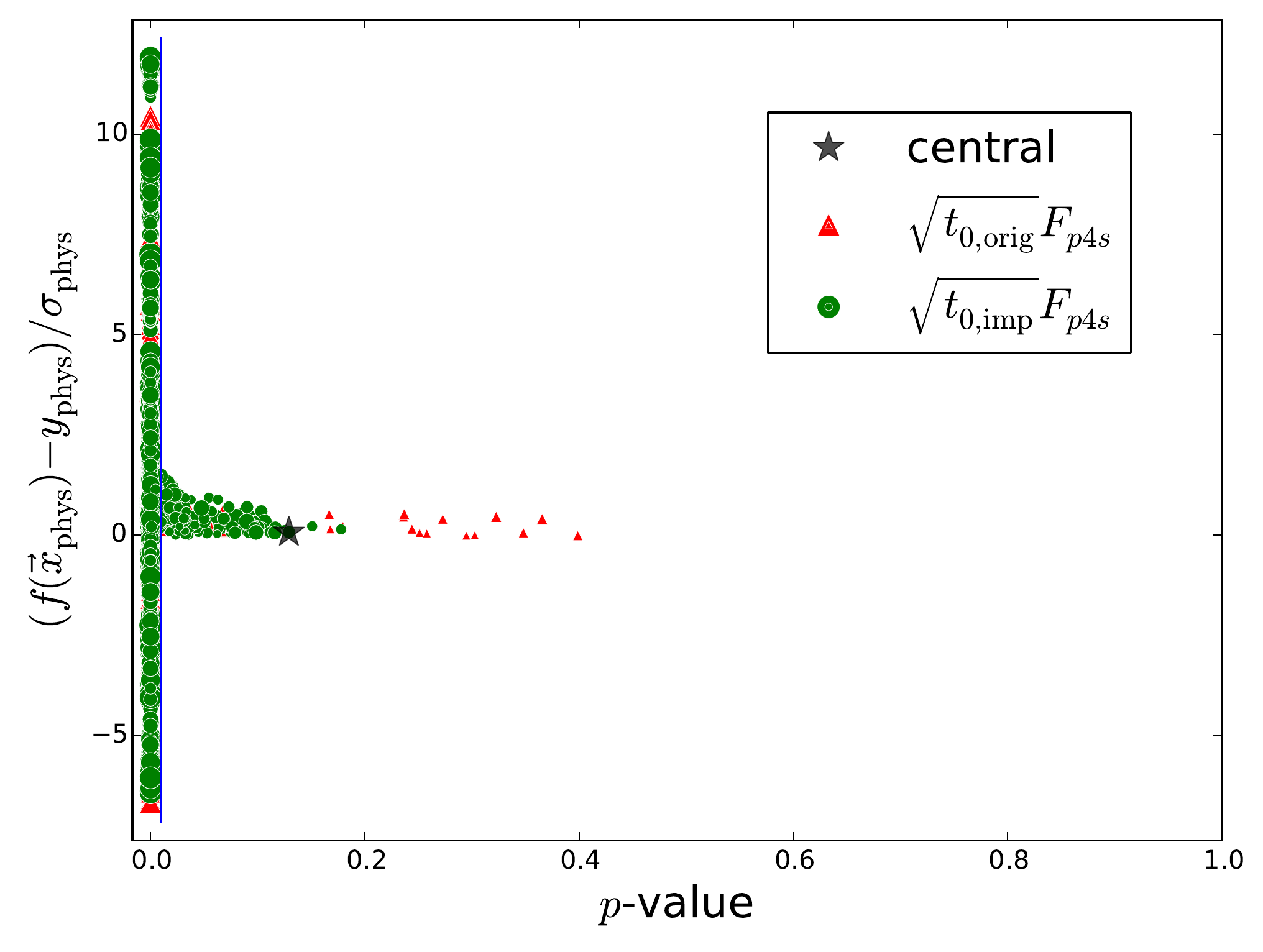}
\includegraphics[scale=0.58]{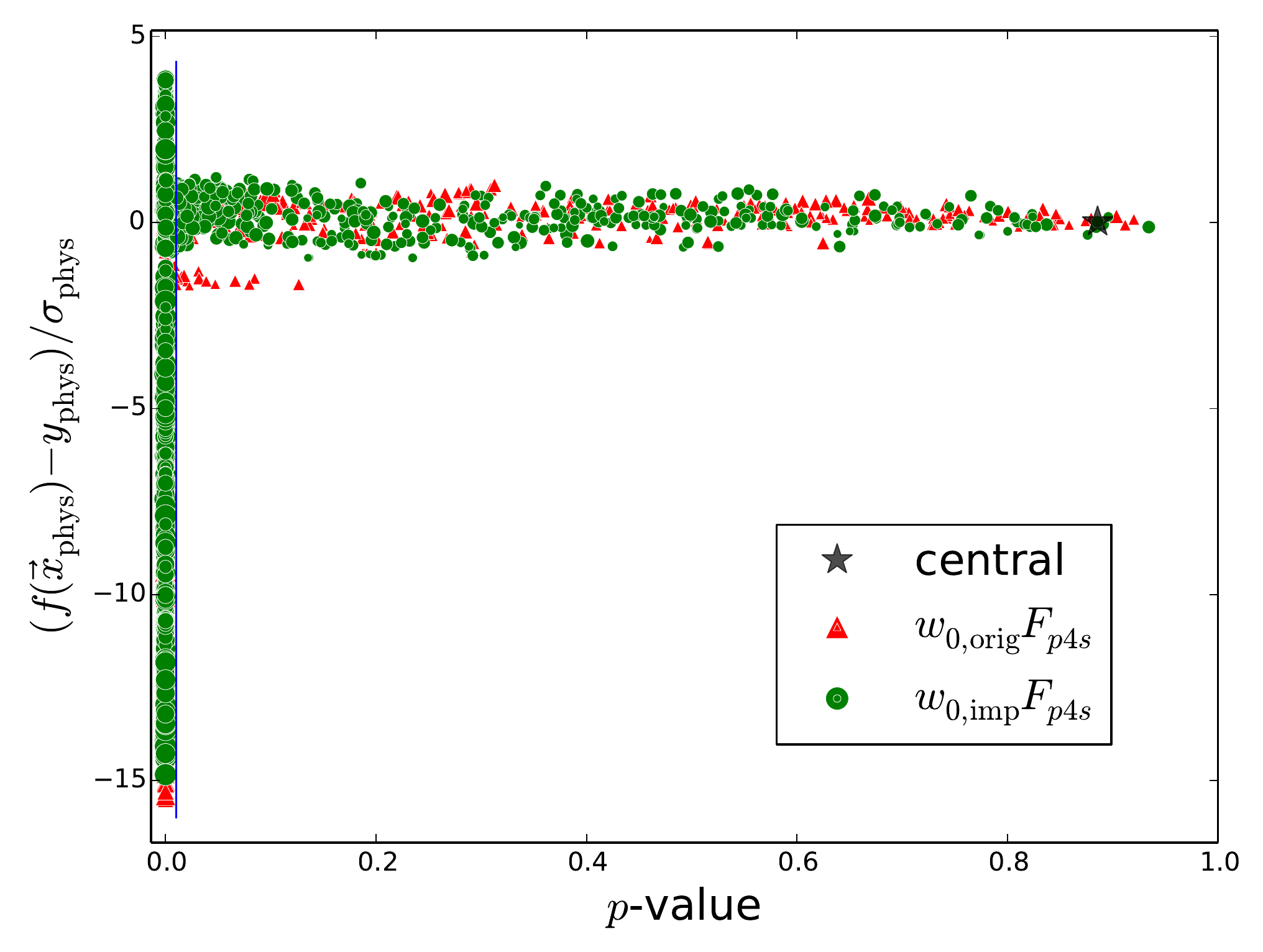}
\caption{\label{fig:acceptfits} 
The ``acceptability" for the various fits considered for the $t_0$ scales ($\sqrt{t_{0,{\rm orig}}}$ and $\sqrt{t_{0,{\rm imp}}}$, top panel) and $w_0$ scales ($w_{0,{\rm orig}}$ and $w_{0,{\rm imp}}$, bottom panel). 
Fit acceptability is determined by the $p$ value ($x$ axis) and further illustrated by the proximity to the results from the physical-mass $a\approx0.06$ fm ensemble in units of $\sigma_{stat}$ ($y$ axis). 
The size of the points is proportional to the number of degrees of freedom.
The region to the right of the black line contains fits with $0.01<p<1.0$ and a deviation of less than $2\sigma_{\rm stat}$. 
This line determines the acceptable subset of fits considered in the subsequent analysis. 
The central fit chosen from this analysis is denoted by the star.}
\end{figure}

We gauge the  acceptability of each of the fits outlined in Sec.~\ref{fitforms} using the $p$ value of the fit. 
We also consider the number of degrees of freedom for each fit and the proximity of the fit curve to the data from our most important ensemble, the one with physical quark masses and $a\approx0.06$ fm.
This extra information is not used to restrict the set of fits, but allows us to better visualize their properties.
Figure \ref{fig:acceptfits} shows the acceptability for the original and improved scales with the $p$ value as the $x$ axis, deviation from the physical $a\approx0.06$ fm ensemble as the $y$ axis, and the size (radius) of each data point proportional to the number of degrees of freedom. 
We define ``acceptable" fits as those with $p>0.01$. 
Acceptable fits are those to the right of the black line in \figref{acceptfits}. 
Note that, for all the scales considered,  fits with acceptable $p$ values are usually close to the result from the $a\approx0.06$ fm physical-mass ensemble. 
For all the gradient-flow scales, no acceptable fit deviates from that result by more than $2\sigma_{\rm stat}$.

\begin{figure}
\includegraphics[scale=0.6]{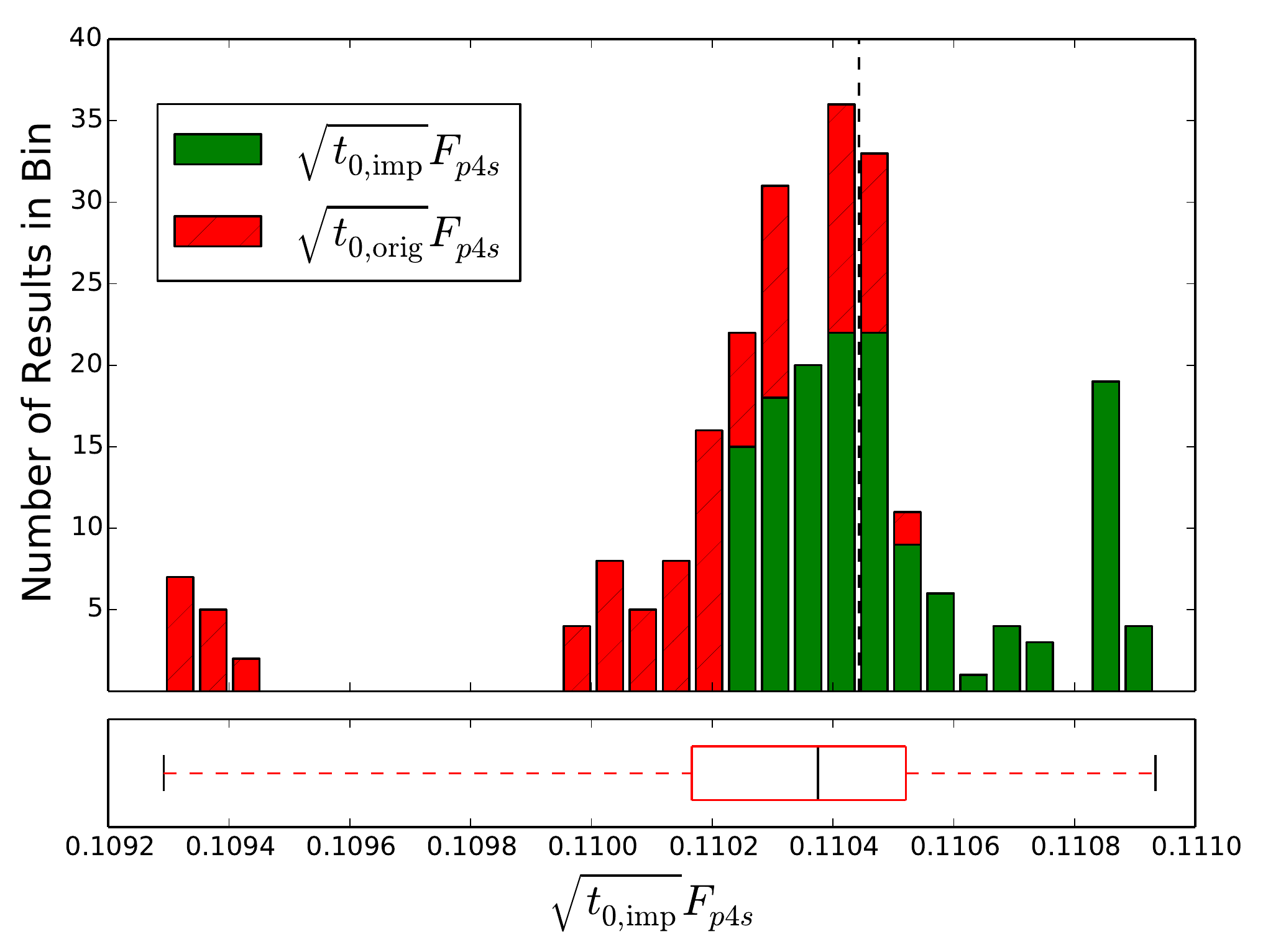}
\includegraphics[scale=0.6]{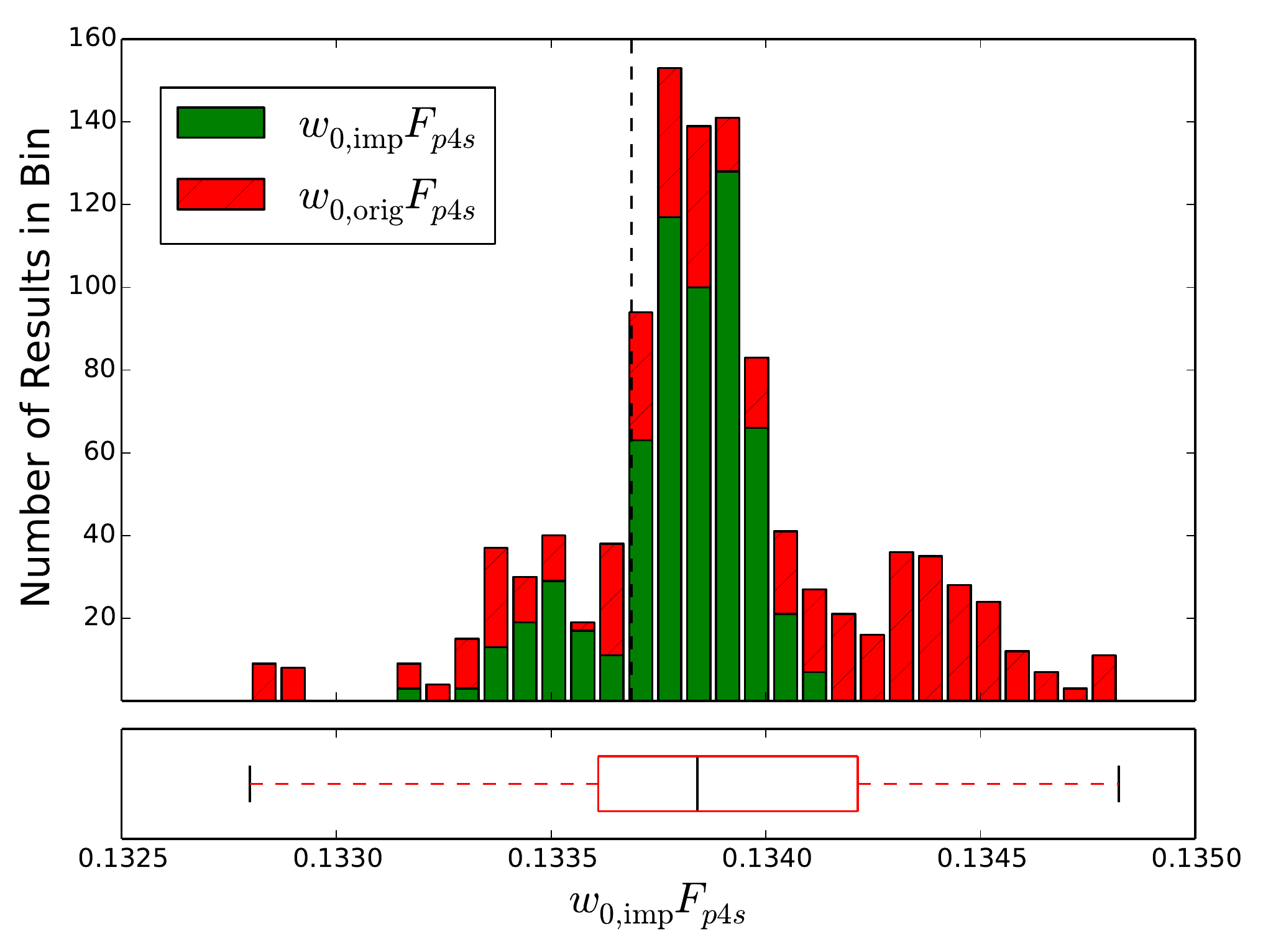}
\caption{\label{fig:histogram} Histograms of the continuum extrapolations for $\sqrt{t_0}F_{p4s}$ (top panel) and $w_0 F_{p4s}$ (bottom panel) for all acceptable fits (see the text). 
Each histogram is a stacked combination of continuum extrapolations from the original ($\sqrt{t_{0,{\rm orig}}}$ and $w_{0,{\rm orig}}$) and improved scales ($\sqrt{t_{0,{\rm imp}}}$ and $w_{0,{\rm imp}}$), represented by the red, hashed and green, solid bars, respectively. 
The box and error bars along the bottom denote the minimum, mean, maximum, and central 68\% of the distribution. 
The vertical dashed line for each distribution marks the continuum result of the associated central fit.}
\end{figure}	

To determine a central value and systematic error from the choice of fit we construct histograms in \figref{histogram} of the continuum results from fits with $p>0.01$. 
A histogram method to determine systematic errors has been used previously by the BMW Collaboration \cite{bmwhist}.  
A key distinction is that we do not treat the distribution as a kind of probability distribution, but simply treat all acceptable fits as realistic alternatives and take the largest positive and negative differences from the central fit as the systematic errors.
For $\sqrt{t_0}$ where the tree-level improvement produces a clear reduction in discretization errors (as discussed later in this section), we use only the histogram of the improved scales $\sqrt{t_{0,{\rm imp}}}$ in the estimate of the systematic error.
In other words, we use the full range of the fits shown in green in \figref{histogram}, but do not consider the red outliers at the left of the histogram.
For $w_0$ the tree-level improvement does not clearly reduce the size of discretization effects, so we include $w_0$ as well as $w_{0,{\rm imp}}$  in the systematic error estimate.
Widening the prior on the NLO charm-mass dependence makes a noticeable shift in the continuum values of some of the outlying fits on both histograms. 
However, widening the prior does not significantly change the continuum values for the central bulk of the histogram. 
For this reason, and because the widest prior width for NLO charm-mass corrections included in the histogram is large ($\sim 1.5\%$, more than three times what is estimated in \rcite{mc_corr}), we do not widen the prior further.

\begin{figure}
\includegraphics[scale=0.6]{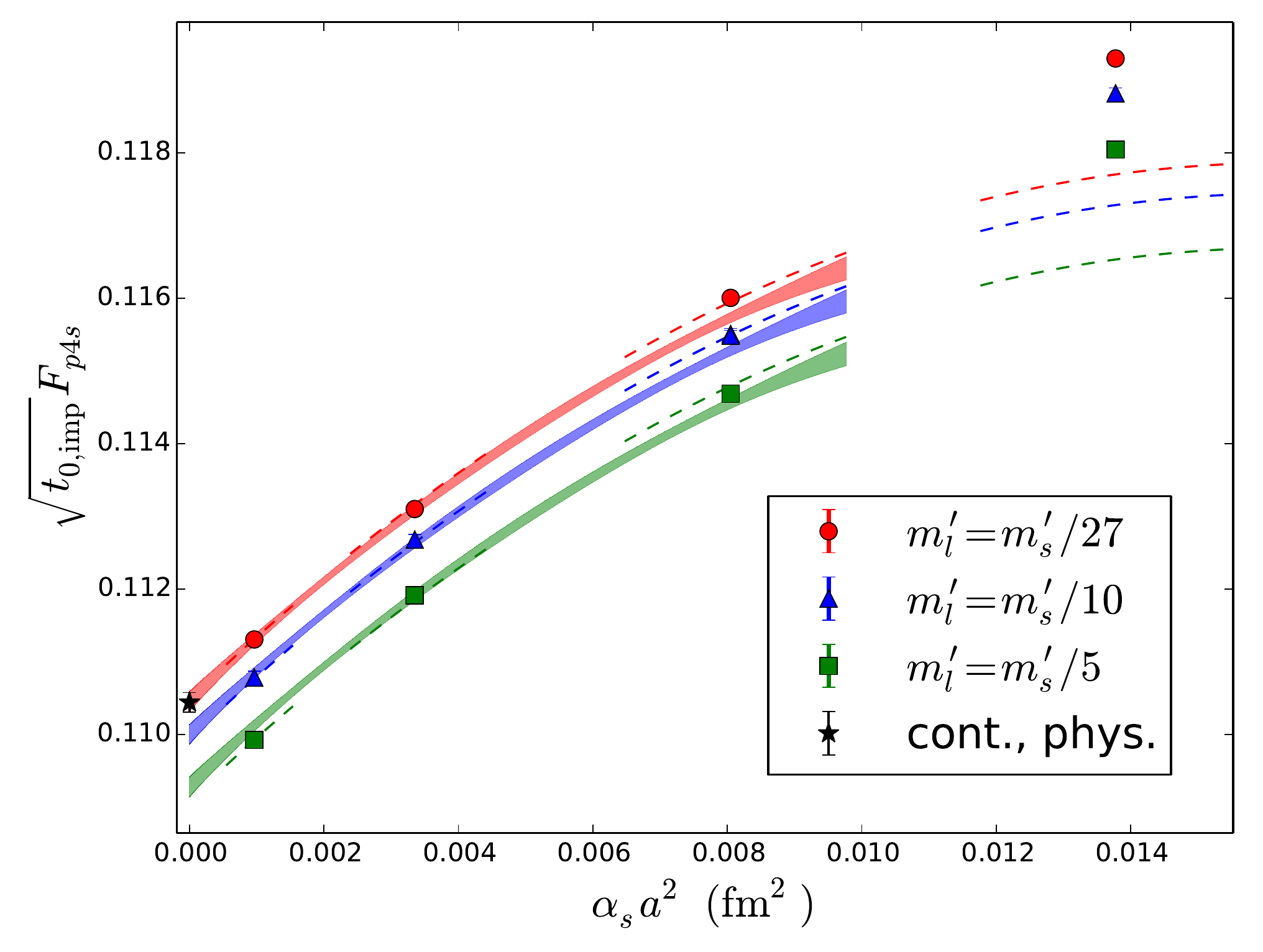}
\includegraphics[scale=0.6]{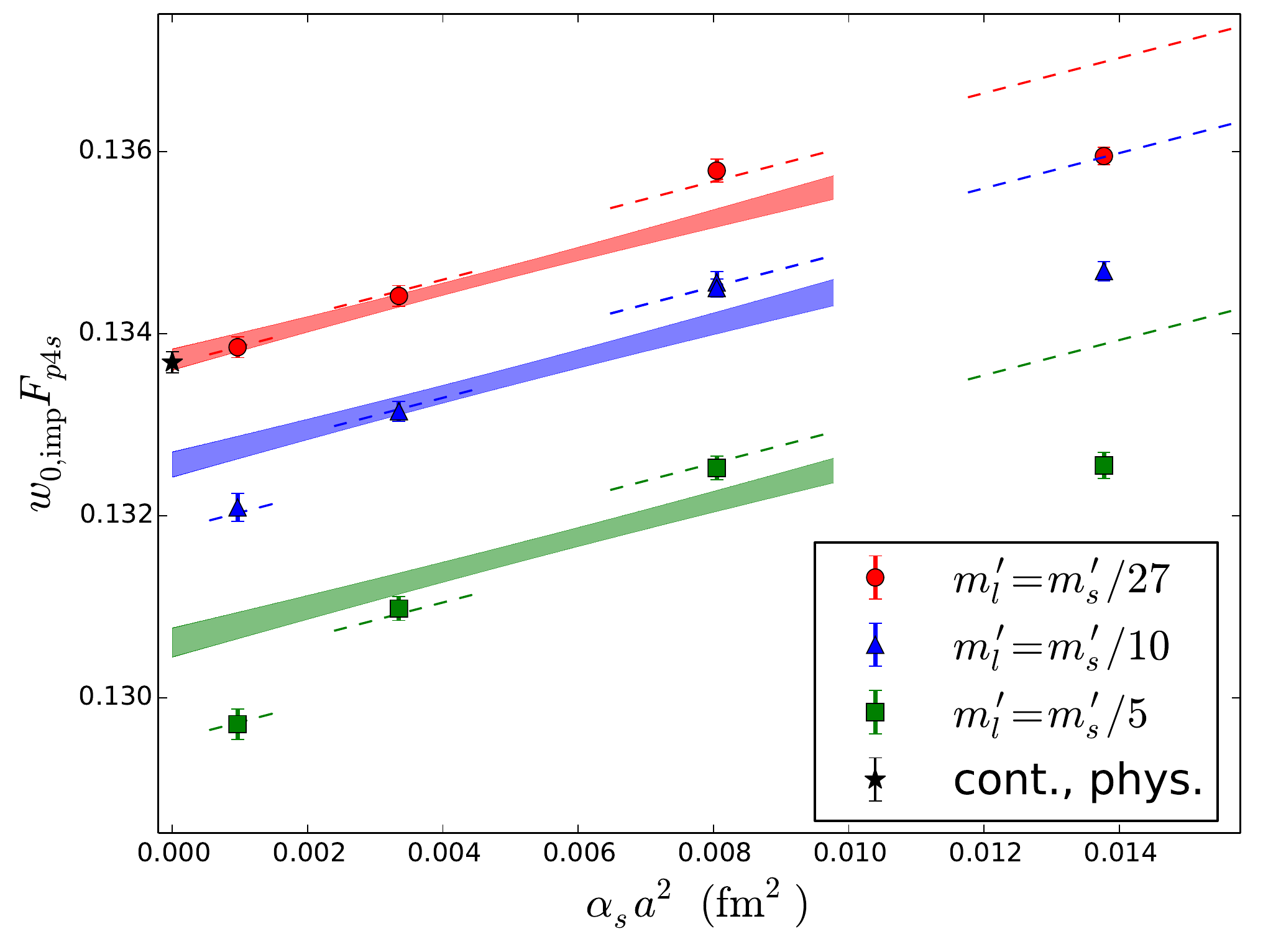}
\caption{\label{fig:centerfit} 
The central fits to the gradient-flow scales $\sqrt{t_{0,{\rm imp}}}F_{p4s}$ and $w_{0,{\rm imp}}F_{p4s}$, plotted as a function of $\alpha_sa^2$. 
These are used to compute $\sqrt{t_0}$ (top panel) and $w_0$ (bottom panel) at physical quark masses and in the continuum, as indicated by the black stars. 
Only $m'_s\!\approx\!m_s$ ensembles are plotted, but the fits include $0.25m_s < m'_s \le m_s$ ensembles. 
Dashed lines represent the fit through each ensemble's actual quark masses and lattice spacing, while the solid bands are for varying lattice spacing at fixed quark masses retuned to the physical strange-quark mass and the ratio of $m_l'/m_s'$ specified in the legend.}
\end{figure}

For both $\sqrt{t_0}$ and $w_0$, the central fit is chosen by locating fits close to the median and mean with $p>0.1$. 
If there are several fits that satisfy this criterion, fits with a  larger number of degrees of freedom are chosen.
For $\sqrt{t_0}$, where there are very few fits with $p>0.1$, this criterion is sufficient to determine the central fit.
For $w_0$, there are a large number of fits satisfying this criterion; thus, we narrow the choice down by preferring fits with $\alpha_sa^2$ over $a^4$, $1/m_c'^2-1/m_c^2$ over $1/m_c'-1/m_c$, and the NNLO chiral expansion over the NLO expansion.
The central fits to $\sqrt{t_0}$ and $w_0$ are both to the improved scales, include all but the three lightest $m_s'$ ensembles, use the $1/m_c'^2-1/m_c^2$ NLO correction in $m_c$, include the $\alpha_sa^2$ lattice-spacing term, exclude the coarsest $a\approx 0.15$ fm ensembles, and use the wider set of priors.
The central fit for $\sqrt{t_0}$ only uses the chiral expansion to NLO and adds in the $a^4$ lattice-spacing term, resulting in five free parameters (with four priors---continuum values are never constrained by priors) and 14 data points.
The central fit for $w_0$ includes the full NNLO chiral expansion but does not add in additional lattice-spacing terms, resulting in seven free parameters (with six priors) and 14 data points.
For $\sqrt{t_0}$, the central fit has $\chi^2/{\rm d.o.f} = 14.0/9$ and $p=0.14$ from the data alone ({\it i.e.}, with the standard or ``unaugmented'' definition of $\chi^2$ coming from data, and degrees of freedom equal to the number of data points minus the number of fit parameters).  Including contributions from priors, the ``augmented'' $\chi^2/{\rm d.o.f} =15.2/13$ and $p=0.31$.  The fit is $0.07\sigma$ higher than the result on the physical $a\approx0.06$ fm ensemble. 
For $w_0$, the central fit has $\chi^2/{\rm d.o.f} = 3.0/7$, $p=0.89$ unaugmented and $\chi^2/{\rm d.o.f} = 4.0/13$, $p=0.99$ augmented,  and is $0.15\sigma$ higher than the result on the physical $a\approx0.06$ fm ensemble.
The central fits are shown in \figref{centerfit}. 
The dashed lines indicate how well the fit describes the data by showing the fit function evaluated at the same masses and lattice spacing as the data points. 
The three solid bands show the lattice-spacing dependence at fixed quark masses, tuned to a physical value for the strange-quark mass and the indicated ratio of the light-quark to strange-quark mass. 
One clearly sees the effects of retuning the quark masses from their simulation values.

\begin{figure}
\includegraphics[scale=0.6]{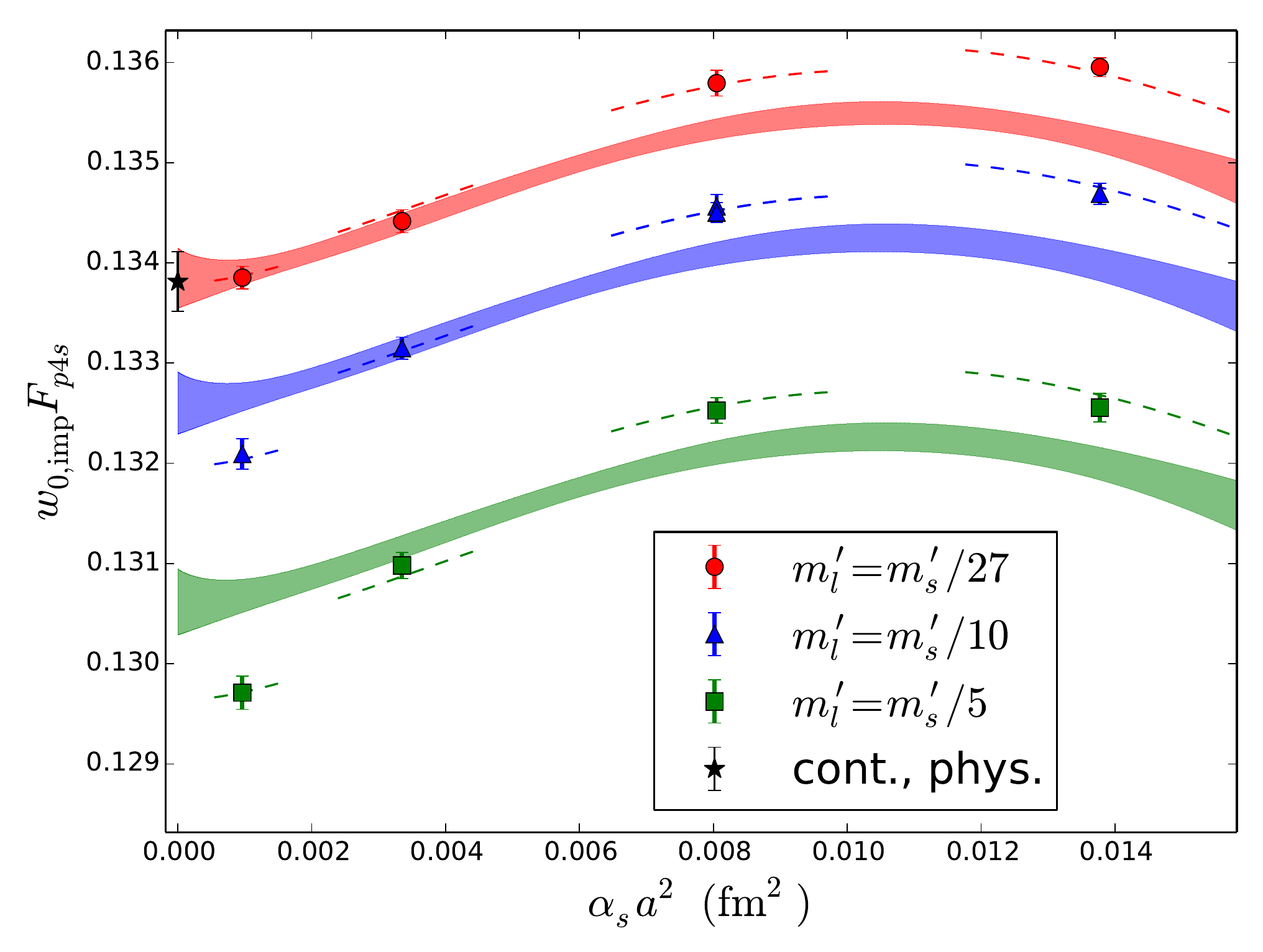}
\caption{\label{fig:alternative} 
An alternative fit to the gradient-flow scale $w_{0,{\rm imp}}F_{p4s}$, plotted as a function of $\alpha_sa^2$. 
The value of $w_0$ at physical quark masses and in the continuum estimated from this fit is indicated by the black star. 
Only $m'_s\!\approx\!m_s$ ensembles are plotted, but the fit includes $0.25m_s <m'_s \le m_s$ ensembles. 
Dashed lines represent the fit through each ensemble's actual quark masses and lattice spacing, while the solid bands are for varying lattice spacing at fixed quark masses retuned to the physical strange-quark mass and the ratio of $m_l'/m_s'$ specified in the legend.}
\end{figure}

Because there are a wide range of choices for the cental fit to $w_0$, we include a description of an alternative fit, which is plotted in \figref{alternative}. 
The fit is similar to the central fit for $w_0$ previously mentioned, except it includes the lattice-spacing terms $a^4$ and $\alpha_s^2a^2$ and the coarsest ensembles at $a\approx0.15$ fm.
The fit has $\chi^2/{\rm d.o.f} = 7.7/8$, $p=0.47$ unaugmented and $\chi^2/{\rm d.o.f} = 9.64/16$, $p=0.89$ augmented, and is $0.18\sigma$ higher than the physical $a\approx0.06$ fm ensemble.
The addition of more lattice-spacing terms and the coarsest ensembles leads to a hook in the continuum extrapolation near $a\approx0.06$ fm, which significantly increases the statistical error in the continuum result.
For this reason we prefer the previously mentioned central fit for $w_0$ over this alternative.

For the fit to $\sqrt{t_{0,{\rm imp}}}F_{p4s}$, the lattice-spacing dependence at finer lattice spacings ($a\le0.09$ fm) is dominated by the $\alpha_sa^2$ contribution. 
The $a^4$ contributions start to become comparable to those from $\alpha_sa^2$ for $a\gtwid0.12\;$ and produce the curvature evident in \figref{centerfit} (top panel).  
The chosen central fit to $w_{0,{\rm imp}}F_{p4s}$ has only one lattice-spacing dependent term, $\alpha_s a^2$, but it excludes the coarsest $a\approx 0.15$ fm ensembles.
So, it is also worthwhile to examine the alternative fit's lattice-spacing dependence since the $a\approx0.15$ fm ensembles are included in this fit.
The lattice-spacing dependence of $w_{0,{\rm imp}}F_{p4s}$ in the alternative fit is milder than for $\sqrt{t_{0,{\rm imp}}}F_{p4s}$, but also more complicated.  
The majority contribution for all $a$ is from $\alpha_s^2a^2$, but only by a slim margin;
the $\alpha_s^2a^2$ term is approximately $53\%-57\%$ of the sum of the absolute value of all discretization terms.
The contributions from $\alpha_s a^2$ and $a^4$ have comparable magnitudes and add together, canceling some of the contribution from $\alpha_s^2a^2$.
The cancellations are larger for $a\le0.06$ fm and $a>0.12$ fm, causing the curvature seen in \figref{alternative}.
Because the central fit to $\sqrt{t_{0,{\rm imp}}}F_{p4s}$ and the alternative fit to $w_{0,{\rm imp}}F_{p4s}$ have multiple discretization terms of comparable magnitudes for $a>0.12$fm, we ensure higher-order terms are negligible by repeating these fits with the addition of the next highest terms in $a^2$ or $\alpha_sa^2$.
The continuum results for these modified fits do not significantly differ from the original fits.

\begin{figure}
\includegraphics[scale=0.7]{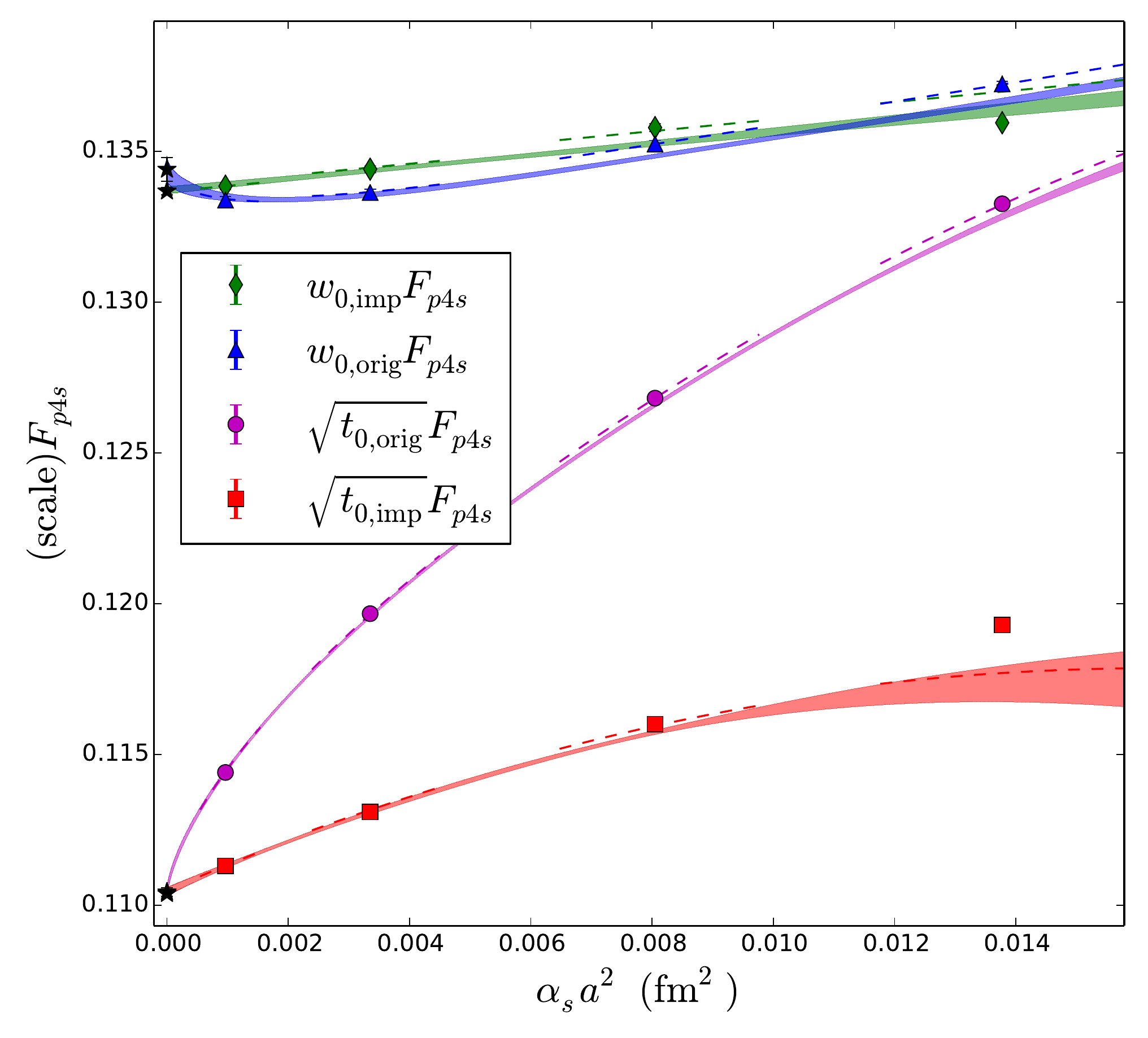}
\caption{\label{fig:scalecomparison} 
Continuum extrapolations for the original ($\sqrt{t_{0,{\rm orig}}}$ and $w_{0,{\rm orig}}$) and improved ($\sqrt{t_{0,{\rm imp}}}$ and $w_{0,{\rm imp}}$) gradient-flow scale times $F_{p4s}$ plotted for physical quark-mass ensembles only. 
All fits to the original, unimproved scales include the chiral expansion to NNLO, $1/m_c^2$ NLO charm-quark corrections, the wider set of priors, all four lattice spacings, and all but the three lightest $m_s'$ ensembles. 
For $\sqrt{t_{0,{\rm orig}}}F_{p4s}$ the fit is quadratic in $a^2$. 
For $w_{0,{\rm orig}}F_{p4s}$ the fit is linear in $a^2$ and $\alpha_s a^2$. 
For the improved scales the plotted lines are from the central fits discussed in this section. 
The continuum-extrapolation points are shown in black with error bars representing the statistical error only.}
\end{figure}

It is revealing to examine the central extrapolations plotted through only the physical-mass ensembles for all four gradient-flow scales, as was done in \figref{naiveImp} for the naive extrapolation.
This plot is presented in \figref{scalecomparison}.
Compared to the simpler fits to just the physical-mass ensembles in Sec.~\ref{naive}, quark-mass mistunings in the physical quark-mass ensembles are accounted for here. 
This leads the two coarsest physical-mass ensembles ($a=0.12$ and $a=0.15$ fm) to shift down when retuned to the precise ratio $m_l/m_s=1/27$.
For the fits to $\sqrt{t_{0,{\rm orig}}}F_{p4s}$ and $\sqrt{t_{0,{\rm imp}}}F_{p4s}$ the difference is visible but has only a small effect on the continuum extrapolation. 
For the fits to $w_{0,{\rm orig}}F_{p4s}$ and $w_{0,{\rm imp}}F_{p4s}$ the shift is very important, as the fluctuation in the data points across the range of $a^2$ is comparable to the size of the effect of the mass retuning. 

For both $\sqrt{t_0}$ and $w_0$, the tree-level improved version of each scale eliminates $a^2$ errors (but not $\alpha_s a^2$) and reduces $a^4$ and $a^6$ contributions. 
The improvement in $\sqrt{t_0}F_{p4s}$ is obvious in \figref{scalecomparison}.  
For $w_0$, even after quark-mass retuning, the size of discretization effects for $a\ltwid 0.06$ fm in $w_{0,{\rm imp}}F_{p4s}$ is at best marginally smaller than in $w_{0,{\rm orig}}F_{p4s}$.
Although  numerical results cannot separate the effects of $F_{p4s}$ from $w_0$, the lack of clear improvement between $w_0$ and $w_{0,{\rm imp}}$ suggests that the dominant lattice artifacts in $w_0$ may not arise at tree level. 
Alternatively, the lattice artifacts from $F_{p4s}$ may be dominating the continuum extrapolation, making it difficult to resolve the differences between $w_{0,{\rm orig}}$ and $w_{0,{\rm imp}}$.

\section{Results}

\subsection{Scales in physical units \label{cont}}
We compute our final estimate of the gradient-flow scales in physical units by evaluating the continuum-extrapolated, physical-quark-mass-interpolated value of $\sqrt{t_0}F_{p4s}$ and $w_0F_{p4s}$ for the best fit in Sec.~\ref{fitresults} and dividing by the physical value of $F_{p4s}$ (see Sec.~\ref{hisq}).
\begin{eqnarray}
\sqrt{t_0} &=& 0.1416 (2)_{stat} ({}^{+6}_{-3})_{t_0,extrap} ({}^{+3}_{-2})_{F_{p4s},extrap} (2)_{\rm FV} (3)_{f_{\pi}\,{\rm PDG}} \ {\rm fm} \eqn{t0_result}\\
      w_0  &=& 0.1714 (2)_{stat} ({}^{+15}_{-11})_{w_0,extrap} ({}^{+3}_{-2})_{F_{p4s},extrap} (2)_{\rm FV} (3)_{f_{\pi}\,{\rm PDG}}\  {\rm fm} \eqn{w0_result}
\end{eqnarray}

The first error is statistical and is from the corresponding central fit discussed in Sec.~\ref{fitresults}.  
The remaining, systematic, errors are from continuum extrapolation/chiral interpolation (estimated by variations among fits),  corresponding continuum and chiral errors on $F_{p4s}$ in physical units, residual finite-volume effects on $F_{p4s}$, and the error in $F_{p4s}$ from the experimental error in $f_\pi$ \cite{PDG}, respectively. 
The error from the choice of fit for the gradient-flow scale is estimated using the histograms in  \figref{histogram}. 
We use the full range of fits to $t_{0,{\rm imp}}F_{p4s}$ for $t_0$ and the full range of all fits for $w_0$.
The remaining extrapolation errors, residual finite-volume effects, and error from the experimental value of $f_\pi$ come directly from the analysis of $F_{p4s}$ \cite{fp4s}.

The results in \eqs{t0_result}{w0_result} may be compared to the earlier, simple estimates of $\sqrt{t_0} = 0.1419(1)({}_{-\phantom{1}4}^{+17})$ fm and $w_0 = 0.1710(4)({}_{-12}^{+\phantom{1}7})$ fm from the physical quark-mass ensembles in Sec.~\ref{naive}.
For both $\sqrt{t_0}$ and $w_0$, the extrapolated values agree, within the earlier systematic errors.  
[Note that the earlier result did not include the uncertainties from $F_{p4s}$ and $f_\pi$, which give the last three errors in  \eqs{t0_result}{w0_result}.] 
For $\sqrt{t_0}$, the central value from the simpler fit is slightly higher and both extrapolations lead to similar statistical uncertainties. 
The main improvement of extrapolating $\sqrt{t_{0,{\rm imp}}}F_{p4s}$ over the full set of ensembles is the narrower systematic uncertainty in the continuum, physical-mass extrapolation. 
For $w_0$, the central value from the simpler fit is slightly lower. 
This shift is attributable to the quark-mass retuning and higher-order discretization terms only accessible to the full extrapolation.  
This additional systematic control leads us to prefer this full analysis over the simple one, even though the total errors for $w_0F_{p4s}$ are slightly larger in the full analysis.
Overall, the addition of nonphysical quark-mass ensembles reduces some uncertainties and improves control over the continuum extrapolation without significantly deviating from our initial estimate.

The results presented in this work have evolved from preliminary results presented previously. 
In chronological order, the estimates from  two earlier proceedings are $w_0=0.1711(2)(8)$ fm in \rcite{prelim1}, and $\sqrt{t_0}=0.1422(2)(5)$ fm and $w_0=0.1732(4)(8)$ fm in \rcite{prelim2}. 
For comparison to the results in this work, we have altered the original results by keeping only the statistical and systematic error from the choice of fit form to $\sqrt{t_0}F_{p4s}$ or $w_0F_{p4s}$. 
We have dropped all other systematic errors, which are shared across all results. 
For both scales, all results agree within 2$\sigma$ of the current results. 
Compared to the  result in \rcite{prelim1}, those in \rcite{prelim2} account for leading-order charm-quark-mass mistunings, use $aF_{p4s}$, instead of $af_\pi$, to set the scale, and consider a larger selection of discretization terms.  
However \rcite{prelim2} uses an incorrect value of $am_c$ for the physical quark mass, $a\approx 0.06$ fm ensemble when adjusting for charm-quark-mass mistunings.
The mistake is fixed in the current work and is responsible for most of the downward shift relative to the scales presented in \rcite{prelim2}.
We have also updated the statistical errors from $aF_{p4s}$ and now include the
induced correlations from $aF_{4ps}$  among ensembles at the same $\beta$.  
Finally, compared to \rcite{prelim2}, the current work incorporates the tree-level improved versions of each scale, refines the selection of discretization terms, includes next-to-leading-order charm-quark-mass corrections, and uses priors to constrain the fit parameters.

\subsection{Continuum meson-mass dependence \label{mass}}

Using the best fits and data sets chosen in Sec.~\ref{fitresults}, we determine the continuum meson-mass dependence of $w_0$ under a mass-independent scale-setting scheme. 
The resulting function is useful for a prediction of the scales on future ensembles, as well as for explicit comparison of the mass dependence of $w_0$ to that of other scale-setting quantities. 
To predict a scale one measures $w_{0,{\rm orig}}/a$ (or $w_{0,{\rm imp}}/a$), $aM_\pi$, and $aM_K$ on a subset of the ensemble to be generated. 
Then, by evaluating the function at the corresponding dimensionless variables $P=(w_0 M_\pi)^2$ and $K=(w_0 M_K)^2$ one can determine the continuum value of $w_0$ in physical units at those masses, $w_0(P,K)$, and compute the resulting scale $a= w_0(P,K)/(w_0/a)$. 
This procedure was originally suggested in \rcite{bmw}.

\begin{figure}
\includegraphics[scale=0.60]{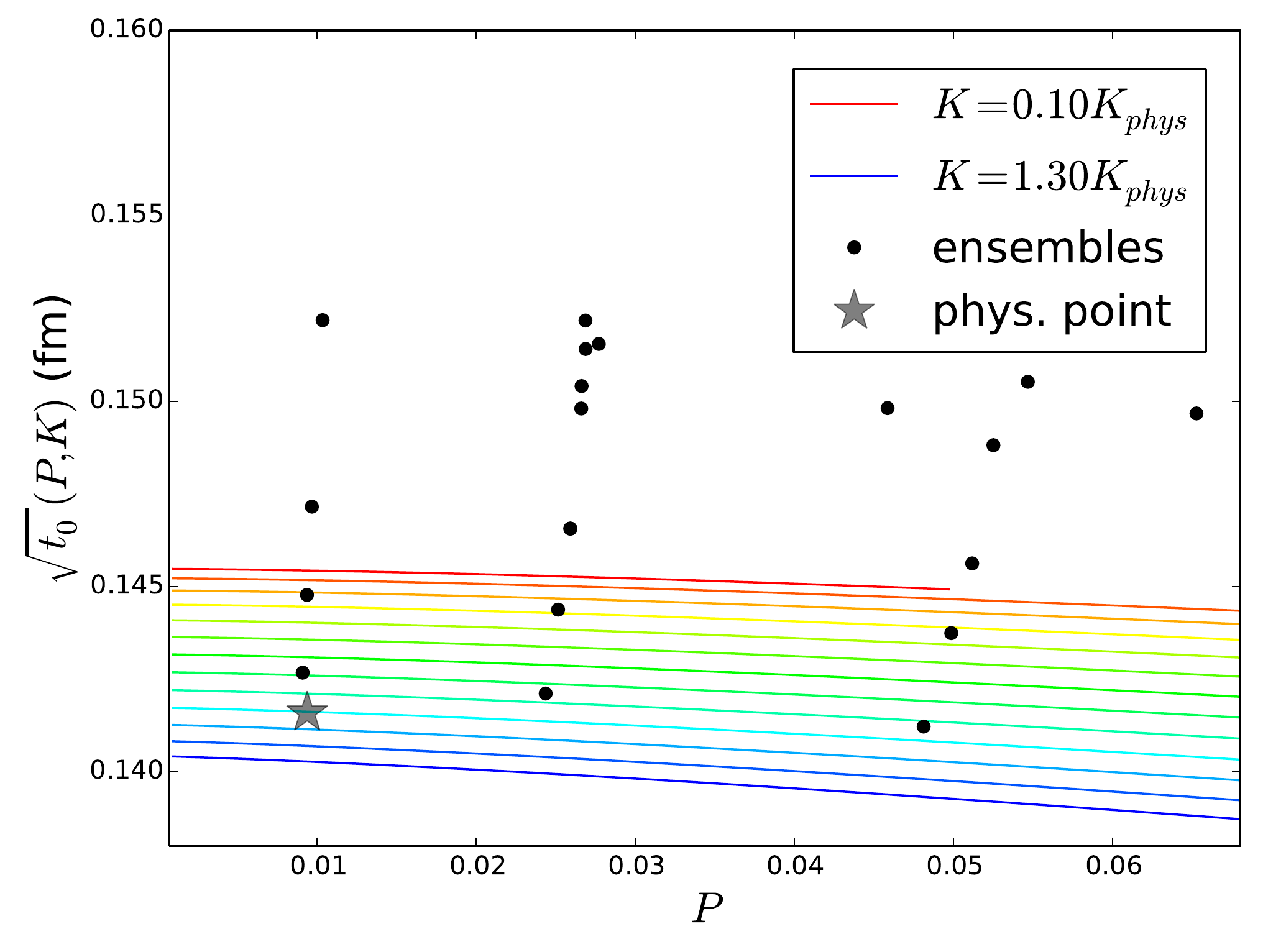}
\includegraphics[scale=0.60]{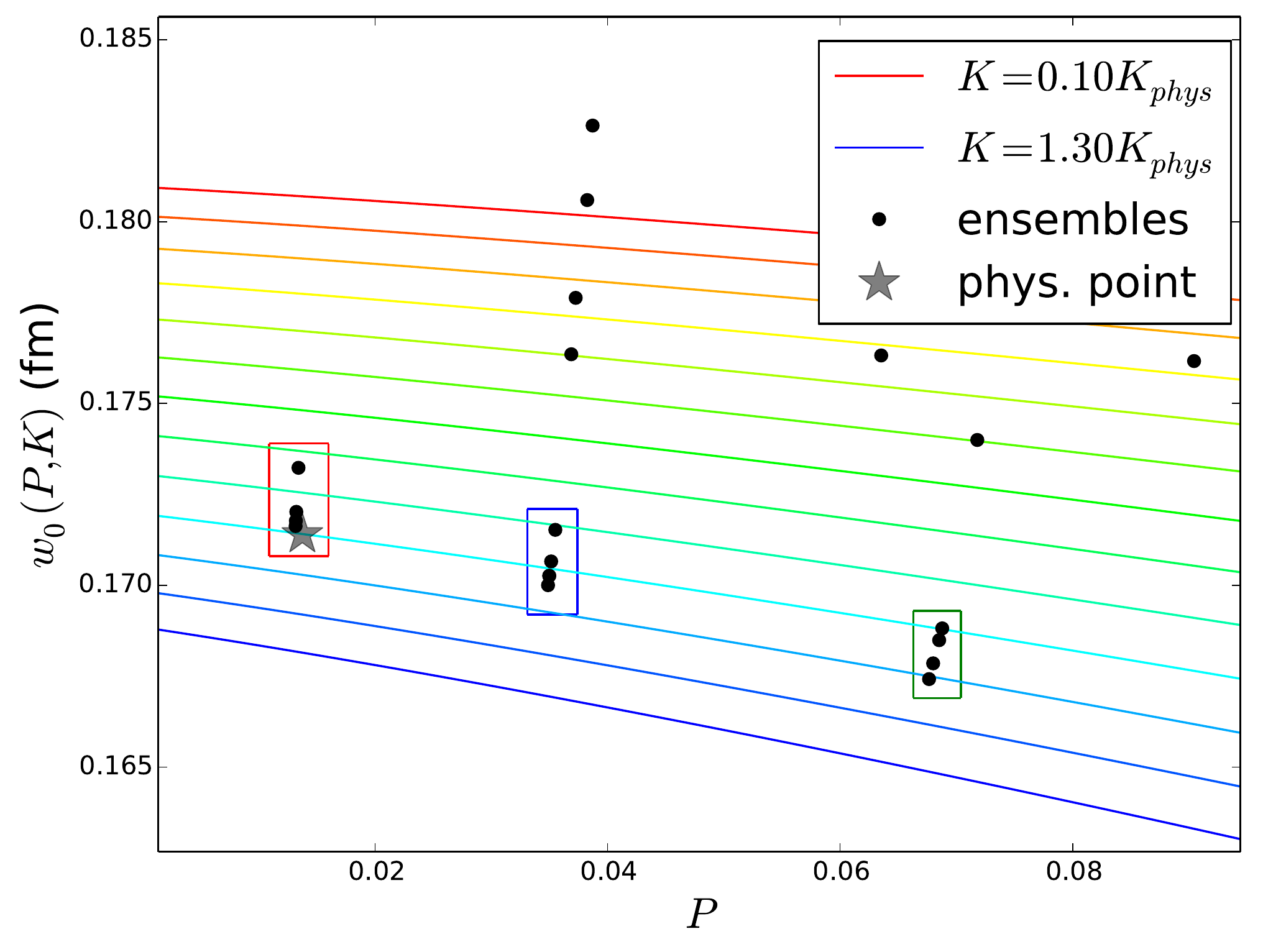}
\vspace{-6mm}
\caption{\label{massPlot} 
The continuum mass dependence of $\sqrt{t_0}$ and $w_0$ as a function of  $P=(w_0 M_\pi)^2$ for fixed values of $K=(w_0 M_K)^2$. 
The black points and the star illustrate the values of the pion and kaon masses that correspond to various HISQ ensembles and to the physical point, respectively.
The three boxes on the plot of $w_0$ enclose the physical strange mass ensembles with different ratios of $m_l'/m_s'$. 
From the left- to the rightmost box the ratios are $m_l'/m_s' = 1/27$, $1/10$, and $1/5$.
Similar boxes are not drawn for $\sqrt{t_0}$ due to the large discretization effects separating the points that would go in each box.
The lines corresponding to $K=0.1K_{\rm phys}$ do not extend over the full domain of $P$ because we restrict $(w_0M_\eta)^2 \approx (4K^2-P^2)/3 \geq 0$.}
\end{figure}

The functional form of the meson-mass dependence $w_0(P,K)$ is chosen to be the same as the chiral expansion to NNLO, in agreement with the best fit chosen in Sec.~\ref{fitresults}. 
The coefficients are determined by solving the implicit equation
\begin{equation}
w_0=f(P=(w_0M_\pi)^2,K=(w_0M_K)^2)
\end{equation}
numerically for $w_0=w_0(P,K)$. Using the best fit 
$h(a, (M_\pi/F_{p4s})^2, (M_K/F_{p4s})^2)=w_0F_{p4s}$ of Sec.~\ref{fitresults}, the implicit function is defined as 
\begin{equation}
w_0(P,K) = h(0, P/(w_0F_{p4s})^2, K/(w_0F_{p4s})^2)/F_{p4s}\ ,
\end{equation}
where $F_{p4s}$ is evaluated at physical quark masses and in the continuum. 
Note, the first parameter is set to 0, denoting the continuum limit. 
We find
\begin{eqnarray} \eqn{massfit_w}
	w_0(P,K) & = & 0.1809 - 0.0055\,(2K+P) + 0.0766\,P\mu_P + 0.0948\,K\mu_K \\
			 & - & 0.0018\,(P-4K)\mu_\eta + 0.0237\,\eta\,\mu_\eta - 0.0363\,(2K+P)^2 + 0.0063\,(K-P)^2 \nonumber
\end{eqnarray} 

where $\mu_z = z\log(z/\Lambda)$, with $\Lambda = (M_\rho/\sqrt{8}\pi F_{p4s})^2 \approx 0.3170$, and $\eta=(4K-P)/3$. 
The fractional error in $w_0(P,K)$ is approximately the same as for our continuum determination of $w_0$ at physical masses, given in \eq{w0_result}. 
Figure \ref{massPlot} plots this function over a large range of values of $P$ and $K$. 
Values corresponding to the HISQ ensembles and to the physical-mass point are overlaid to give a sense of the range of meson masses for which this function is valid. 
The leading $(2K+P)$ dependence can be observed in the roughly linear shape for each line of constant $K$ and the approximately constant vertical gap between lines of fixed $K$, independent of $P$.   
The separation of points within the clusters of physical strange-quark mass ensembles that were simulated close to the physical ratios $m_l/m_s=1/5,1/10,$ and $1/27$ is due to quark-mass mistunings and discretization errors. 

Using \eq{massfit_w} and the results for $w_{0,{\rm imp}}/a$ on the HISQ ensembles, we recalculate $a$(fm) for each ensemble and check to see that the results are consistent with the original lattice spacings set through $F_{p4s}$. 
\tabref{new_scales} lists the lattice spacings determined through $F_{p4s}$ in \rcite{fp4s} and $w_0$ in this work. 
The scales determined from $w_{0,{\rm imp}}$ are almost independent of quark masses for fixed $\beta$, showing that the procedure is working as designed, and can be used to find consistent scales of new ensembles, even if they do not have physical quark masses. 
Lattice spacings determined from $F_{p4s}$ and $w_{0,{\rm imp}}$ on the physical quark-mass ensembles agree as the continuum limit is approached, and are close over the whole range of lattice spacings. 
This fitting procedure may be repeated to find $\sqrt{t_0}$ as a function of $P'=(\sqrt{t_0} M_\pi)^2$ and $K'=(\sqrt{t_0} M_K)^2$.
Because the central fit for $\sqrt{t_0}F_{p4s}$ does not include NNLO terms from chiral-perturbation theory, we redo the fit with NNLO terms added. 
We find
\begin{eqnarray} \eqn{massfit_t}
	\sqrt{t_0}(P,K) & = & 0.1455 + 0.0007(2K+P) + 0.0994\,P\mu_P + 0.1336\,K\mu_K \\
	         & + & 0.0002\,(P-4K)\mu_\eta + 0.0334\,\eta\,\mu_\eta - 0.0214\,(2K+P)^2 - 0.0040\,(K-P)^2, \nonumber
\end{eqnarray}
\noindent
where the notation and error determination are the same as for $w_0$, and the fractional error of $\sqrt{t_0}(P,K)$ is approximately that of $t_0$ in \eq{t0_result}.
The corresponding mass-dependence and lattice-spacing estimates are shown in Fig.~\ref{massPlot} and \tabref{new_scales}.
As might be expected from the large slope seen for $\sqrt{t_0}$ in \figref{naiveImp}, the lattice-spacing estimates show large discretization effects for the coarser ensembles.

\begin{table}
\caption{\label{tab:new_scales} 
Values of the lattice spacing determined from $aF_{p4s}$\cite{fp4s}, $w_{0,{\rm imp}}/a$, and $\sqrt{t_{0,{\rm imp}}}/a$ on the physical strange-quark HISQ ensembles listed in \tabref{hisqPhys}. 
The first two columns list the coupling $\beta$ and the ratio of light- to strange-sea-quark masses, with the lattice dimensions appended as needed to uniquely identify each ensemble. 
Since we do not have a function corresponding to \eqs{massfit_w}{massfit_t} for $F_{p4s}$, a mass-independent scale setting with $F_{p4s}$ is performed on the physical quark-mass ensembles only. 
The error listed with each estimate of $a$ from the gradient-flow scales comes from the full error of $w_0(P,K)$ or $\sqrt{t_0}(P,K)$.
Errors from quark-mass mistunings are not included; in other words we are finding the scale at the actual simulation values of the quark masses, rather than at the intended values.}
\begin{tabular}{ccccc}
\hline\hline
$\beta$ & $m_l'/m_s'$            & \ $(aF_{p4s})/F_{p4s}$(fm)  \ & \ $w_0/(w_{0,{\rm imp}}/a)$(fm)\ & \ $\sqrt{t_0}/(\sqrt{t_{0,{\rm imp}}}/a)$(fm)\  \\
\hline
\ $5.80$\   & $1/5$                  & \ldots                         & $0.1511 ({}^{+18}_{-15})$ & $0.1410 ({}^{+15}_{-12})$ \\
$5.80$  & $1/10$                 & \ldots                         & $0.1511 ({}^{+14}_{-11})$ 	& $0.1413 ({}^{+\phantom{1}9}_{-\phantom{1}6})$ \\
$5.80$  & $1/27$                 & $0.15305({}^{+57}_{-41})$ & $0.1509 ({}^{+14}_{-11})$ 		& $0.1413 ({}^{+\phantom{1}9}_{-\phantom{1}6})$ \\
\hline
$6.00$  & $1/5$                  & \ldots                         & $0.1206 ({}^{+14}_{-12})$ 	& $0.1162 ({}^{+11}_{-\phantom{1}9})$ \\
$6.00$  & \ $1/10\ (32^3\times64)$\  & \ldots                         & $0.1206 ({}^{+11}_{-\phantom{1}9})$ 	& $0.1163({}^{+8}_{-5})$ \\
$6.00$  & $1/10\ (40^3\times64)$ & \ldots                         & $0.1207 ({}^{+11}_{-\phantom{1}9})$ 		& $0.1163 ({}^{+8}_{-5})$ \\
$6.00$  & $1/27$                 & $0.12232({}^{+45}_{-33})$ & $0.1206 ({}^{+11}_{-\phantom{1}9})$ 		& $0.1163 ({}^{+7}_{-5})$ \\
\hline
$6.30$  & $1/5$                  & \ldots                         & $0.0873 ({}^{+11}_{-\phantom{1}9})$ 	& $0.0855 ({}^{+8}_{-7})$ \\
$6.30$  & $1/10$                 & \ldots                         & $0.0874 ({}^{+8}_{-7})$ 	& $0.0857 ({}^{+6}_{-4})$ \\
$6.30$  & $1/27$                 & $0.08791({}^{+33}_{-24})$ & $0.0875 ({}^{+8}_{-7})$ 	& $0.0858 ({}^{+5}_{-3})$ \\
\hline
$6.72$  & $1/5$                  & \ldots                         & $0.0566 ({}^{+7}_{-6})$ 	& $0.0561 ({}^{+6}_{-5})$ \\
$6.72$  & $1/10$                 & \ldots                         & $0.0565 ({}^{+6}_{-5})$ 	& $0.0561 ({}^{+4}_{-3})$ \\
$6.72$  & $1/27$                 & $0.05672({}^{+21}_{-16})$ & $0.0566 ({}^{+5}_{-4})$ 	& $0.0562 ({}^{+3}_{-2})$ \\
\hline\hline
\end{tabular}
\end{table}

\section{Discussion and Conclusions \label{end}}

\begin{figure}
\includegraphics[scale=0.6]{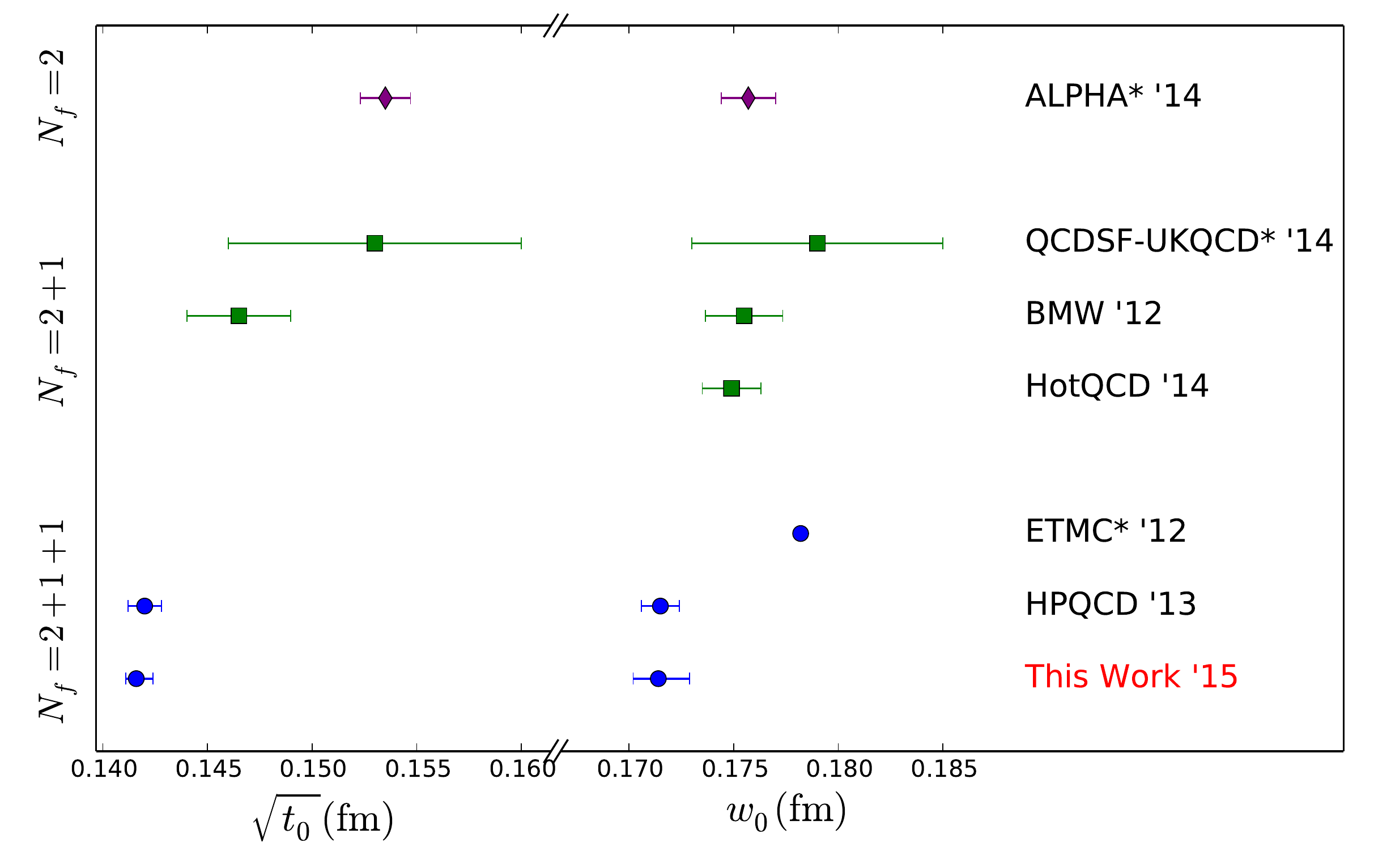}
\caption{\label{fig:collabPlot} 
The continuum values of $\sqrt{t_0}$ and $w_0$ separated by collaboration and grouped by the number of flavors. 
References for each collaboration's work can be found in \tabref{contComp}. 
Those results for collaborations marked with an asterisk are preliminary. 
Our results for $\sqrt{t_0}$ and $w_0$ are consistent within 2 standard deviations to all other results except the preliminary calculations from the ALPHA Collaboration.}
\end{figure}

With the continuum results complete, we compare with computations of gradient-flow scales performed by other collaborations. 
Table \ref{tab:contComp} shows a selection of those calculations and their final results in comparison with our own. 
The same results are also plotted in \figref{collabPlot}. 
Differences are shown divided by the joint error, except for the HPQCD Collaboration data. 
Because HPQCD uses a subset of the HISQ ensembles employed here, we do not use the joint sigma, which would double count several sources of error; instead, we use the larger of the two collaborations' total error. 
Our results for both scales are compatible with those of the three other published continuum-limit calculations by HPQCD, HotQCD, and BMW; the largest difference is 1.9$\sigma$. 
Our best agreement is with HPQCD, the latter of which performed an independent analysis on the same HISQ configurations but without the $a = 0.06$~fm ensembles. 
We also agree with the published, single-lattice-spacing result for $\sqrt{t_0}=0.1414(7)(5)$ fm from TWQCD \cite{twqcd}.
Furthermore, we agree within $2\sigma$ with all but one collaboration's preliminary results: $\sqrt{t_0}$ and $w_0$ calculated by the ALPHA Collaboration with $N_f=2$. 
This may be due to the difference in the number of flavors: \Rcite{alpha} has found stronger $N_f$ dependence for $\sqrt{t_0}$ than for $w_0$, which is consistent with the observed deviations between the ALPHA Collaboration's preliminary results and those of this paper\cite{alpha}.

\begin{table}
\caption{\label{tab:contComp} 
Continuum results for the gradient-flow scales $\sqrt{t_0}$ and $w_0$ from different collaborations \cite{hpqcd,etm,bmw,alpha,ukqcd,hotqcd}.
The last two columns tabulate the difference between the results of other collaborations and this work, relative to one joint sigma. 
For HPQCD, whose errors are not independent of ours, we simply use the larger error for the comparison. 
Results of the three collaborations marked with an asterisk are the preliminary conference results. }
\begin{tabular}{cccccc}
\hline\hline
Collaboration & $N_f$ & $\sqrt{t_0}$ (fm) & $\Delta \sqrt{t_0} / \sigma$ & $w_0$ (fm) & $\Delta w_0 / \sigma$ \\
\hline
MILC [This work]          & \ 2+1+1 \   & \ $0.1416 (1)({}^{+8}_{-5})$ \   & \ldots     & \ $0.1714 (2)({}^{+15}_{-12})$\   & \ldots     \\
HPQCD        \cite{hpqcd} & 2+1+1   & $0.1420 (8)$                 & $+0.4$ & $0.1715 (9)$                            & $+0.1$ \\
ETMC*        \cite{etm}   & 2+1+1   & \ldots                           & \ldots     & $0.1782$                                & \ldots     \\
HotQCD       \cite{hotqcd}& 2+1     & \ldots                           & \ldots     & $0.1749 (14)$ & $+1.8$ \\
BMW          \cite{bmw}   & 2+1     & $0.1465 (21)(13)$            & $+1.9$ & $0.1755 (18)(04)$                       & $+1.7$ \\
QCDSF-UKQCD* \cite{ukqcd} & 2+1     & $0.153  (7)$                 & $+1.6$ & $0.179  (6)$                            & $+1.2$ \\
ALPHA*       \cite{alpha} & 2       & $0.1535 (12)$                & \ $+8.3$\  & $0.1757 (13)$                           & \ $+2.2$ \ \\
\hline\hline
\end{tabular}
\end{table}

Finally, we compare the relative lattice scale found from $\sqrt{t_{0,{\rm orig}}}$, $w_{0,{\rm orig}}$, and  other quantities used for scale setting. 
Here, we assume that the scale setting is being performed on ensembles with physical quark masses, so that extrapolation in quark mass is not required for any quantity.  
In that case, the systematic errors associated with extracting any of these scales on a given, physical-mass ensemble are generally significantly smaller than the statistical errors,  with the possible exception of $r1/a$ at finer lattice spacings, for which errors in extracting asymptotic energies may become significant.
Table \ref{stats} compares the percent statistical error for various scale-setting quantities in lattice units measured on the HISQ physical quark-mass ensembles. 
Both gradient-flow scales are determined more precisely than $r_1/a$ and $af_\pi$.  
The precision of $\sqrt{t_{0,{\rm orig}}}/a$ is higher than, and the precision of $w_{0,{\rm orig}}/a$ is on par with, the most precise of the other scales, $aF_{p4s}$. 
This small statistical error was an original motivation for computing the scale from gradient flow. 
Note further that $\sqrt{t_{0,{\rm orig}}}/a$ and $w_{0,{\rm orig}}/a$ have only been determined on a small 
subset of the configurations at finer lattice spacings, while the $aF_{p4s}$ values come from the entire ensembles, so there is considerable room for improvement for the gradient-flow scales.
In addition, lower systematic errors---in particular, low dependence on quark masses---may make the gradient-flow scales preferable to $aF_{p4s}$ for relative scale setting, especially when scales are needed for ensembles with unphysical quark masses or with significant quark-mass tuning errors. 
Statistical errors for $w_{0,{\rm orig}}/a$ are larger than those of $\sqrt{t_{0,{\rm orig}}}/a$. 
This is one factor, although not the dominant factor, that leads to our slightly more precise continuum extrapolated value for $\sqrt{t_0}$ compared to $w_0$.
On the other hand, \figref{scalecomparison} illustrates that the discretization effects for $w_{0,{\rm orig}}$ are much smaller than those for $\sqrt{t_{0,{\rm orig}}}$ when compared with the reference scale $aF_{p4s}$.  
It is conceivable that the small slope for $w_{0,{\rm orig}}$  and  $w_{0,{\rm imp}}$ is due to an accidental cancellation between their discretization errors and those of $F_{p4s}$. 
However, when combined with the empirical evidence given in \rcite{bmw}, it appears more likely that $w_0$ has ``intrinsically'' smaller $a^2$ dependence than $\sqrt{t_0}$ in the sense  that the ratio of $w_0$ to most common reference scales will have smaller discretization errors than the corresponding ratio for $\sqrt{t_0}$.
Finally, we remark that the small error of $aF_{p4s}$, in comparison with that of $af_\pi$, is what motivates us to use $aF_{p4s}$ for our continuum extrapolations of the gradient-flow scales, as discussed in Sec.~\ref{naive}.

\begin{table}
\caption{\label{stats} 
Percent statistical error for several scale-setting quantities including $r_1$, $f_\pi$, $F_{p4s}$, and the gradient-flow scales $\sqrt{t_{0,{\rm orig}}}/a$ and $w_{0,{\rm orig}}/a$ on the physical quark-mass HISQ ensembles listed in \tabrefs{hisqPhys}{gfPhys}. 
The statistical errors in the improved scales $\sqrt{t_{0,{\rm imp}}}$ and $w_{0,{\rm imp}}$ are comparable to the original gradient-flow scales. 
The first column is the approximate lattice spacing and can be used to uniquely identify each ensemble.}
\begin{tabular}{cccccc}
\hline\hline
\multirow{2}{*}{$\approx a$(fm)} & \multicolumn{5}{c}{Statistical Error (\%)} \\
 \cline{2-6}
 & \ $r_1/a$\  & \ $af_\pi$\  & \ $aF_{p4s}$\  & \ $\sqrt{t_0}/a$\  & \ $w_0/a$\  \\
\hline
$0.15$ & $0.36$ & $0.11$ & $0.06$ & $0.02$ & $0.03$ \\
$0.12$ & $0.25$ & $0.08$ & $0.06$ & $0.04$ & $0.07$ \\
$0.09$ & $0.33$ & $0.09$ & $0.06$ & $0.03$ & $0.07$ \\
$0.06$ & $0.12$ & $0.11$ & $0.06$ & $0.03$ & $0.06$ \\
\hline\hline
\end{tabular}
\end{table}

In conclusion, we have computed the continuum, physical-mass values of $\sqrt{t_0}$ and $w_0$, and find $\sqrt{t_0}=0.1416({}^{+8}_{-5})$ fm and $w_0=0.1714({}^{+15}_{-12})$ fm, in reasonable agreement with most independent calculations, and in excellent agreement with the results of HPQCD, who used a subset of the same HISQ ensembles employed here. 
We have estimated the integrated autocorrelation lengths at different lattice spacings and found autocorrelation lengths comparable to that of the topological charge, although the errors at the finer lattice spacing ($a\approx 0.09$fm) are quite large.
Compared to our preliminary work, the continuum extrapolation here is better controlled through the removal of tree-level discretization errors, the use of $aF_{p4s}$ over $af_\pi$ to set the scale, and the use of priors to suppress outlying fits that have unreasonable lattice-spacing or charm-mass dependence. 
Further, the quark-mass interpolation has been constrained using chiral-perturbation theory, and the effect of charm-mass mistunings have been taken into account up to next-to-leading order.  
Finally, we have calculated the continuum meson-mass dependence for use in future scale-setting applications.

\begin{acknowledgments}
Computations for this work were carried out with resources provided by the USQCD Collaboration, the Argonne Leadership Computing Facility and the National Energy Research Scientific Computing Center, which are funded by the Office of Science of the U.S. Department of Energy; and with resources provided by the National Center for Atmospheric Research, the National Center for Supercomputing Applications, the National Institute for Computational Science, and the Texas Advanced Computing Center, which are funded through the National Science Foundation’s Teragrid/XSEDE Program; and with resources provided by the Blue Waters Computing Project, which is funded by NSF Grants No. OCI-0725070 and No. ACI-1238993 and the State of Illinois. This work is also part of the “Lattice QCD on Blue Waters” PRAC allocation supported by National Science Foundation Grant No. OCI-0832315. This work was supported in part by the U.S. Department of Energy under Grants No. DE-FG02-91ER40628 (C.B., N.B., J.K.), No. DE-FC02-12ER41879 (C.D., J.F., L.L.), No. DE-FG02-91ER40661 (S.G.), No. DE-SC0010120 (S.G.), No. DE-FG02-13ER-41976 (D. T.), by the National Science Foundation under Grants No. PHY-1067881 (C. D., J. F., L. L.), No. PHY-10034278 (C. D.),  No. PHYS-1417805 (J. L.), and No. PHY-1316748 (R. S.). This manuscript has been coauthored by an employee of Brookhaven Science Associates, LLC, under Contract No. DE-AC02-98CH10886 with the U.S. Department of Energy. Fermilab is operated by Fermi Research Alliance, LLC, under Contract No. DE-AC02-07CH11359 with the U.S. Department of Energy.
\end{acknowledgments}

\bibliography{gradFlow_HISQ}

\end{document}